\documentclass[preprint]{aastex62}
\usepackage{amsmath,amssymb,CJK,soul}
\usepackage{longtable}
\usepackage{hyperref}
\usepackage{longtable}
\usepackage{threeparttablex} 

\newcommand{\cd}{{2020 CD$_3$ }}
\newcommand{\cdns}{{2020 CD$_3$}}

\received{--}
\revised{--}
\accepted{--}
\submitjournal{ApJL}

\shorttitle{Spectrophotometry of Minimoon \cdns}
\shortauthors{Bolin et al.}

\begin{document}

\title{Characterization of Temporarily-Captured Minimoon \cd by Keck Time-resolved Spectrophotometry}

\correspondingauthor{Bryce Bolin}
\email{bbolin@caltech.edu}

\author[0000-0002-4950-6323]{Bryce T. Bolin}
\affiliation{Division of Physics, Mathematics and Astronomy, California Institute of Technology, Pasadena, CA 91125, U.S.A.}
\affiliation{IPAC, Mail Code 100-22, Caltech, 1200 E. California Blvd., Pasadena, CA 91125, USA}

\author{Christoffer Fremling}
\affiliation{Division of Physics, Mathematics and Astronomy, California Institute of Technology, Pasadena, CA 91125, U.S.A.}

\author[0000-0003-0437-3296]{Timothy R. Holt}
\affiliation{University of Southern Queensland, Computational Engineering and Science Research Centre, Queensland, Australia}
\affiliation{Southwest Research Institute, Department of Space Studies, Boulder, CO-80302, USA}

\author{Matthew J. Hankins}
\affiliation{Division of Physics, Mathematics and Astronomy, California Institute of Technology, Pasadena, CA 91125, U.S.A.}

\author[0000-0002-2184-6430]{Tom{\'a}s Ahumada}
\affiliation{Department of Astronomy, University of Maryland, College Park, MD 20742, USA}

\author{Shreya Anand}
\affil{Division of Physics, Mathematics and Astronomy, California Institute of Technology, Pasadena, CA 91125, USA}


\author{Varun Bhalerao}
\affiliation{Department of Physics, Indian Institute of Technology Bombay, Powai, Mumbai-400076, India}

\author{Kevin B. Burdge}
\affiliation{Division of Physics, Mathematics and Astronomy, California Institute of Technology, Pasadena, CA 91125, U.S.A.}

\author{Chris M. Copperwheat}
\affiliation{Astrophysics Research Institute Liverpool John Moores University, 146 Brownlow Hill, Liverpool L3 5RF, United Kingdom}

\author{Michael Coughlin}
\affil{School of Physics and Astronomy, University of Minnesota, Minneapolis, Minnesota 55455, 
USA}

\author[0000-0001-5253-3480]{Kunal P. Deshmukh}
\affiliation{Department of Metallurgical Engineering and Materials Science, Indian Institute of Technology Bombay, Powai, Mumbai-400076, India}

\author{Kishalay De}
\affiliation{Division of Physics, Mathematics and Astronomy, California Institute of Technology, Pasadena, CA 91125, U.S.A.}

\author{Mansi M. Kasliwal}
\affiliation{Division of Physics, Mathematics and Astronomy, California Institute of Technology, Pasadena, CA 91125, U.S.A.}

\author{Alessandro Morbidelli}
\affiliation{Universit\'{e} C\^{o}te d'Azur, Observatoire de la C\^{o}te d'Azur, CNRS, Laboratoire Lagrange, Boulevard de l'Observatoire, CS 34229, 06304 Nice cedex 4, France}

\author{Josiah N. Purdum}
\affiliation{Department of Astronomy, San Diego State University, 5500 Campanile Dr, San Diego, CA 92182, U.S.A.}

\author{Robert Quimby}
\affiliation{Department of Astronomy, San Diego State University, 5500 Campanile Dr, San Diego, CA 92182, U.S.A.}
\affiliation{Kavli Institute for the Physics and Mathematics of the Universe (WPI), The University of Tokyo Institutes for Advanced Study, The University of Tokyo, Kashiwa, Chiba 277-8583, Japan}

\author[0000-0002-2668-7248]{Dennis Bodewits}
\affiliation{Physics Department, Leach Science Center, Auburn University, Auburn, AL 36832, U.S.A.}

\author[0000-0003-1656-4540]{Chan-Kao Chang}
\affiliation{Institute of Astronomy, National Central University, 32001, Taiwan}


\author{Wing-Huen Ip}
\affiliation{Institute of Astronomy, National Central University, 32001, Taiwan}



\author{Chen-Yen Hsu}
\affiliation{Institute of Astronomy, National Central University, 32001, Taiwan}

\author[0000-0003-2451-5482]{Russ R. Laher}
\affiliation{IPAC, Mail Code 100-22, Caltech, 1200 E. California Blvd., Pasadena, CA 91125, USA}

\author{Zhong-Yi Lin}
\affiliation{Institute of Astronomy, National Central University, 32001, Taiwan}

\author{Carey M. Lisse}
\affiliation{Johns Hopkins University Applied Physics Laboratory, Laurel, MD 20723}

\author{Frank J. Masci}
\affiliation{IPAC, Mail Code 100-22, Caltech, 1200 E. California Blvd., Pasadena, CA 91125, USA}

\author{Chow-Choong Ngeow}
\affiliation{Institute of Astronomy, National Central University, 32001, Taiwan}


\author{Hanjie Tan}
\affiliation{Institute of Astronomy, National Central University, 32001, Taiwan}

\author{Chengxing Zhai}
\affiliation{Jet Propulsion Laboratory, California Institute of Technology, 4800 Oak Grove Drive, Pasadena, CA 91109, U.S.A.}

\author{Rick Burruss}
\affiliation{Caltech Optical Observatories, California Institute of Technology, Pasadena, CA 91125, U.S.A.}

\author[0000-0002-5884-7867]{Richard Dekany}
\affiliation{Caltech Optical Observatories, California Institute of Technology, Pasadena, CA 91125, U.S.A.}

\author{Alexandre Delacroix}
\affiliation{Caltech Optical Observatories, California Institute of Technology, Pasadena, CA 91125, U.S.A.}

\author[0000-0001-5060-8733]{Dmitry A. Duev}
\affiliation{Division of Physics, Mathematics and Astronomy, California Institute of Technology, Pasadena, CA 91125, USA}

\author{Matthew Graham}
\affiliation{Division of Physics, Mathematics and Astronomy, California Institute of Technology, Pasadena, CA 91125, U.S.A.}

\author{David Hale}
\affiliation{Caltech Optical Observatories, California Institute of Technology, Pasadena, CA 91125, U.S.A.}

\author[0000-0001-5390-8563]{Shrinivas R. Kulkarni}
\affiliation{Division of Physics, Mathematics and Astronomy, California Institute of Technology, Pasadena, CA 91125, U.S.A.}

\author[0000-0002-6540-1484]{Thomas Kupfer}
\affiliation{Kavli Institute for Theoretical Physics, University of California, Santa Barbara, CA 93106, U.S.A.}

\author[0000-0003-2242-0244]{Ashish Mahabal}
\affiliation{Division of Physics, Mathematics and Astronomy, California Institute of Technology, Pasadena, CA 91125, U.S.A.}
\affiliation{Center for Data Driven Discovery, California Institute of Technology, Pasadena, CA 91125, U.S.A.}

\author{Przemyslaw J. Mr\'{o}z}
\affiliation{Division of Physics, Mathematics and Astronomy, California Institute of Technology, Pasadena, CA 91125, U.S.A.}

\author{James D. Neill}
\affiliation{Division of Physics, Mathematics and Astronomy, California Institute of Technology, Pasadena, CA 91125, U.S.A.}

\author[0000-0002-0387-370X]{Reed Riddle}
\affiliation{Caltech Optical Observatories, California Institute of Technology, Pasadena, CA 91125}

\author{Hector Rodriguez}
\affiliation{Caltech Optical Observatories, California Institute of Technology, Pasadena, CA 91125, U.S.A}

\author[0000-0001-7062-9726]{Roger M. Smith}
\affiliation{Caltech Optical Observatories, California Institute of Technology, Pasadena, CA 91125}

\author[0000-0001-6753-1488]{Maayane T. Soumagnac}
\affiliation{Lawrence Berkeley National Laboratory, 1 Cyclotron Road, Berkeley, CA 94720, U.S.A.}
\affiliation{Department of Particle Physics and Astrophysics, Weizmann Institute of Science, Rehovot 76100, Israel}

\author{Richard Walters}
\affiliation{Division of Physics, Mathematics and Astronomy, California Institute of Technology, Pasadena, CA 91125, U.S.A.}

\author{Lin Yan}
\affiliation{Division of Physics, Mathematics and Astronomy, California Institute of Technology, Pasadena, CA 91125, U.S.A.}

\author{Jeffry Zolkower}
\affiliation{Caltech Optical Observatories, California Institute of Technology, Pasadena, CA 91125, U.S.A.}





\begin{abstract}
We present rotationally-averaged visible spectrophotometry of minimoon \cdns, the second asteroid known to become temporarily captured by the Earth-Moon system's gravitational field. The spectrophotometry was taken with Keck I/LRIS  between wavelengths 434 nm and 912 nm in $B$, $g$, $V$, $R$, $I$ and RG850 filters as it was leaving the Earth-Moon system on 2020 March 23 UTC. The broad-band spectrophotometry of \cd most closely resembles the spectra of V-type asteroids and some Lunar rock samples with a reddish slope of $\sim$18$\%$/100 nm between 434 nm and 761 nm corresponding to colors of $g$-$r$ = 0.62$\pm$0.08, $r$-$i$ = 0.21 $\pm$ 0.06 and an absorption band at $\sim$900 nm corresponding to $i$-$z$ = -0.54$\pm$0.10 \citep[][]{DeMeo2009, Isaacson2011}. Combining our measured 31.9$\pm$0.1 absolute magnitude with an albedo of 0.35 typical for V-type asteroids \citep[][]{DeMeo2013}, we determine \cdns's size to be $\sim$1.0$\pm$0.1 m making it the first minimoon and one of the smallest asteroids to be spectrally characterized. We use our time-series photometry to detect periodic lightcurve variations with a $<$10$^{-4}$ false alarm probability corresponding to a lightcurve period of $\sim$573 s and a lightcurve amplitude of $\sim$1 mag implying \cd possesses a $b/a$ axial ratio of $\sim$2.5. In addition, we extend the observational arc of \cd to 37 days between 2020 February 15 UTC and 2020 March 23 UTC. From the improved orbital solution for \cdns, we estimate its likely duration of its capture to be $\sim$2 y, and we measure the non-gravitation perturbation on its orbit due to radiation pressure with an area-to-mass ratio of 6.9$\pm$2.4$\times$10$^{-4}$ m$^2$/kg implying a density of 2.1$\pm$0.7 g/cm$^3$, broadly compatible with the densities of other meter-scale asteroids \citep[e.g.,][]{Micheli2012, Mommert2014} and Lunar rock \citep[$\sim$2.4 g/cm$^3,$][]{Kiefer2012}. We searched for pre-discovery observations of \cd in the ZTF archive as far back as 2018 October \citep[][]{Masci2019}, but were unable to locate any positive detections.
\end{abstract}
\keywords{minor planets, asteroids: individual (\cdns), temporarily captured orbiters, minimoons}

\section{Introduction}
Asteroid population models predict that out of $\sim$10$^9$ asteroids larger than 1 m in the steady-state near-Earth object (NEO) population \citep[][]{Harris2015,SchunovaLilly2017} a small fraction, $\sim$10$^{-7}$, become temporarily captured by the Earth-Moon system's gravity every year \citep[][]{Granvik2012, Jedicke2018}. These temporary natural satellites, or what we call ``minimoons'' have pre-capture orbital trajectories similar to the Earth which allow them to encounter the Earth at relatively low $\sim$1 km/s speeds assisting them in their capture. However, the captured status of the minimoon is temporary due to its interaction with the gravity of the Sun and the Earth-Moon system and other massive Solar System bodies, with the vast majority of minimoons only being gravitationally-bound the Earth-Moon system for $\sim$70- 280 days \citep[][]{Fedorets2017}. Since they originate from the NEO population, the Earth-Moon system has a steady-state of temporarily captured minimoons with 1-2 being in orbit around the Earth at any given time with diameter $\sim$1 m. Out of $\sim$22,000 NEOs currently known\footnote{\texttt{https://minorplanetcenter.net/iau/TheIndex.html}}, only $\sim$5 known NEOs are in the 1 m range. Thus, due to their frequency of capture and small size, minimoons, therefore, provide the opportunity to study the smallest and most incomplete portion of the NEO population \citep[][]{Granvik2018}. Also because of their low velocities relative to the Earth, minimoons provide excellent targets for human exploration missions \citep[][]{Elvis2011,Granvik2013b,Chyba2014}. 

Asteroids on quasi-satellite orbits that are strongly affected by the gravitational influence of the Earth-Moon system but are not gravitationally captured have been observed before, such as 2013 LX28 and (469219) Kamo`oalewa \citep[][]{Sidorenko2014, FuenteMarcos2016}. However, the previously only known example of an asteroid being truly gravitationally captured by the Earth-Moon system was 2006 RH$_{120}$ discovered by the Catalina Sky Survey in 2006 while it was being captured by the Earth-Moon system's gravity. This first known minimoon remained in orbit around the Earth for $\sim$400 days \citep[][]{Kwiatkowski2008} and had a diameter of $\sim$3 m. Recently, the second known example of a temporarily-captured asteroid was discovered on 2020 February 15 UTC with the Catalina Sky Survey's 1.5-meter telescope \citep[][]{Pruyne2020CD3}.

At the time of discovery on 2020 February 15 UTC, \cd had a highly eccentric geocentrically-bound orbit with a geocentric eccentricity, $e_g$, of 0.96, a geocentric semi-major axis, $a_p$, of $\sim$3 Lunar distances (LD) or $\sim$0.008 au, where 1 LD equals $\sim$0.00257 au, and a prograde geocentric inclination $i_g$ of 49$^\circ$. The majority of space-debris or satellites of artificial origin are contained within 0.1 LD of the Earth suggesting that the distant geocentric orbit of \cd is of natural origin \citep[][]{Tingay2013}. It is possible for artificial objects such as spacecraft boosters to be on more distant geocentric orbits that may have trajectories similar to temporary natural orbits, such as J002E3, a possible Apollo program-era rocket booster \citep[][]{Jorgensen2003}. The Minor Planet Center maintains a list of known spacecraft as well as distant space debris\footnote{\texttt{https://minorplanetcenter.net/iau/artsats/artsats.html}}, however none of these known spacecraft or distant space debris were associated with the trajectory of \cd at the time of its discovery \citep[][]{Pruyne2020CD3}.

While the initial orbit suggests a natural origin, determining the origin from within the Solar System is less clear. As discussed above, NEO population models describe that the minimoon population originates from the Main Belt \citep[][]{Granvik2017}, however, another possible natural origin of minimoons is from Lunar impacts due to the typical low, $\sim$1 km ejection speed of Lunar debris \citep[][]{Gladman1995}. The orbits of Lunar debris dynamically decay after a few kyrs, though it is possible that some Lunar ejecta can be re-captured by the Earth-Moon system as minimoons due to their orbital similarity with the Earth just as for minimoons of asteroidal origin \citep[][]{Granvik2012}. However, it is currently unknown from the dynamical circumstances of Lunar debris and temporarily captured asteroids what proportion of minimoons have an origin as the former to the latter.

In this paper, we build on the discovery of \cd with ground-based visible observations of \cd as an observational test to constrain whether \cd is of asteroidal or Lunar debris origin. As we will discuss below, we will use the approach of \citet[][]{Bolin2020asdfasdf} to combine estimation of its taxonomic classification from spectrophotometric observations at different visible wavelengths as well as its physical properties. In addition, we will use the astrometry from our observations of \cd to extend the orbital arc and refine the orbit of \cd enabling study of its dynamical evolution before, during and after its capture by the Earth-Moon system. The refined orbit enabled by our observations will also serve as an independent constraint on its origin before being captured by the Earth-moon system as well as on its physical properties and origin by comparison with the NEO population model \citep[][]{Morbidelli2020} and estimation of non-gravitational perturbations on its orbit \citep[][]{Mommert2014}.

\section{Observations}
\label{s.obs}
 We used the 10 m Keck I telescope with the Low Resolution Imaging Spectrometer (LRIS) \citep[][]{Oke1995,Rockosi2010} to observe \cd on 2020 March 23.545694 UTC to March 23.583322 UTC in imaging mode (Program ID C236, PI M. Fremling). Both the blue camera consisting of a 2 x 2K x 4K Marconi CCD array and the red camera consisting of a science grade Lawrence Berkeley National Laboratory 2K x 4K CCD array were used simultanously. Both cameras have a spatial resolution of 0.135 arcsec/pixel and were used in 2 x 2 binning mode providing an effective resolution element size of 0.27 arcsec providing a field of view of 6\arcmin x 7.8\arcmin. The 560 nm dichroic with $\sim$50$\%$ peak transmission  was used in combination with the filters $B$ ($\lambda_{\mathrm{eff}}$ = 435 nm, FWHM of 91 nm), $g$ ($\lambda_{\mathrm{eff}}$ = 474 nm, FWHM of 98 nm), $V$ ($\lambda_{\mathrm{eff}}$ = 541 nm, FWHM of 95 nm) filters for the blue camera. The filters $R$ ($\lambda_{\mathrm{eff}}$ = 628 nm, FWHM of 119 nm), $I$ ($\lambda_{\mathrm{eff}}$ = 760 nm, FWHM of 123 nm) and RG850  ($\lambda_{\mathrm{eff}}$ = 912 nm, FWHM of 128 nm), similar to the SDSS $z$ filter \citep[$\lambda_{\mathrm{eff}}$ = 905 nm, FWHM of 137 nm,][]{Fukugita1996}, were used for the red camera in total with the blue camera providing six total resolution elements between 435 nm and 912 nm\footnote{\texttt{https://www2.keck.hawaii.edu/inst/lris/filters.html}}. Typical exposure times were $\sim$120 s which were tracked non-siderally at the the $\sim$3\arcsec/min motion of \cdns. We rotated filters and used the two cameras simultaneously to limit the effect of rotational variations on photometric measurements. At the time of our observations, \cd was located near R.A., Dec. = 14 20 00.3, +33 15 49.7, and had a heliocentric distance, $r_h$, of 1.0059 au, a geocentric distance, $\Delta$, of 0.0128 au and a phase angle, $\alpha$, of 45.4063$^\circ$. During our observations, the sky plane of motion of \cd was $\sim$2.5 \arcsec/m and had airmass 1.03-1.08.
 
\begin{figure}
\centering
\includegraphics[scale=.405]{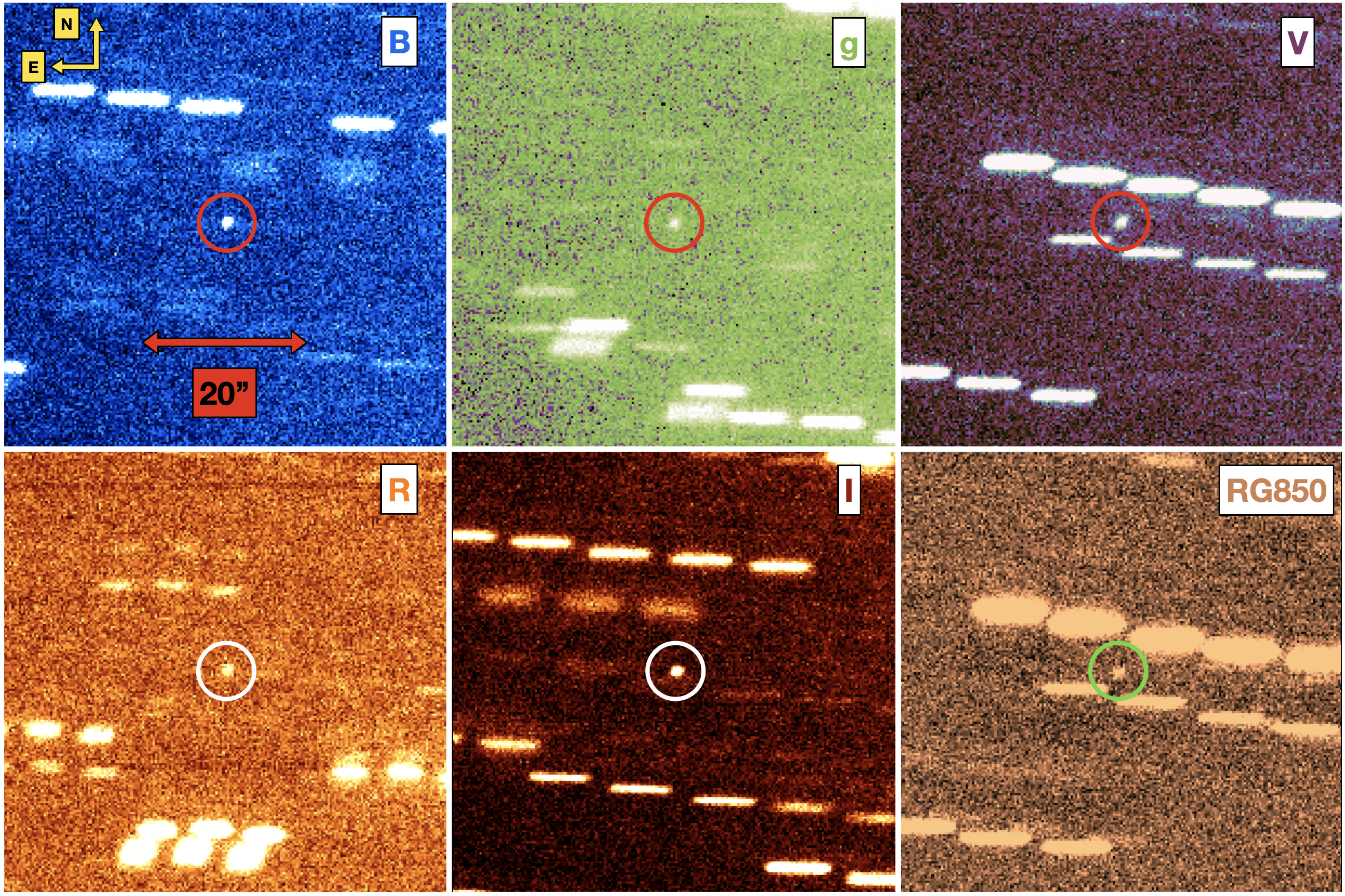}
\caption{Top left panel: a 600 s equivalent exposure time robust mean stack of 5 x 120 s $B$ filter images of \cdns. An arrow indicating the width of 20\arcsec~is shown for scale and the cardinal directions are indicated. Top center panel: a 600 s equivalent exposure time robust mean stack of 4 x 120 s $g$ filter images of \cdns. Top right panel: a 600 s equivalent exposure time robust mean stack of 5 x 120 s $V$ filter images of \cdns. Bottom left panel: a 180 s equivalent exposure time mean stack of 3 x 60 s $R$ filter images of \cdns. Bottom center panel: a 960 s equivalent exposure time robust mean stack of 8 x 120 s $I$ filter images of \cdns. Bottom right panel: a 600 s equivalent exposure time robust mean stack of 5 x 120 s RG850 filter images of \cdns.}
\label{fFiig:keckmosaic}
\end{figure}

 Images were taken of Solar-like calibrator stars in nearby fields as \cd were identified using the Pan-STARRS catalog \citep[][]{Chambers2016,Flewelling2016}. The seeing was $\sim$0.5\arcsec~and data from the CFHT Sky Probe indicated that the night was photometrically stable with less than $\sim$0.01 magnitude variations over the course of our observations\footnote{\texttt{$\mathrm{http://cfht.hawaii.edu/cgi-bin/elixir/skyprobe.pl?plot}\&\mathrm{mcal}\_20200323\mathrm{.png}$}}. Bias and flat frames were obtained using the uniform flattening screen inside the Keck dome. The reduction of the imaging data was completed using the LPipe reduction software \citep[][]{Perley2019}. In total, 5 x 120 s $B$ filter, 5 x 120 s $g$ Filter, 5 x 120 s $V$ filter images, 3 x 30 s, 3 x 60 s $R$ filter images, 8 x 120 s $I$ filter, and 5 x 120 s RG850 filter exposures were taken. Some images contained field stars near \cd which were discarded. The images were median-combined into separate composite stacks for all six filters as shown in the image mosaic in Fig~\ref{fFiig:keckmosaic}.
 
\section{Results}
\subsection{Photometry and spectral classification}
\label{sec:photo}
We measured the photometry of \cd and the Solar analog stars using a 0.81\arcsec~aperture subtracting the median contribution from the sky background within a  2.5-3.5\arcsec~annulus. The $B$,$V$,$R$ and $I$ Johnson-Cousins and $g$ SDSS filter photometry were calibrated using Solar analog stars from the Pan-STARRS catalog \citet[][]{Chambers2016}. The Pan-STARRS catalog magnitudes of the Solar analog stars were transformed to Johnson-Cousins and SDSS magnitudes using the conversions from \citet[][]{Tonry2012}.  We obtain magnitudes $B$ = 25.11 $\pm$ 0.05, $g$ = 24.48 $\pm$ 0.05, $V$ = 24.21 $\pm$ 0.05, $R$ = 23.74 $\pm$ 0.06 and $I$ = 23.31 $\pm$ 0.02. In addition, we determine an RG850 magnitude of 23.88 $\pm$ 0.09 calibrated by using the equivalent SDSS $z$ filter magnitudes determined for our Solar analog from the Pan-STARRS catalog. However, we caution the RG850 and SDSS $z$ filters not being exactly alike, therefore our measured uncertainty is likely affected by small systematic differences between the RG850 and SDSS $z$ filters and thus serves as a lower limit to its true $RG850$ magnitude. 

\begin{figure}
\centering
\includegraphics[scale=.405]{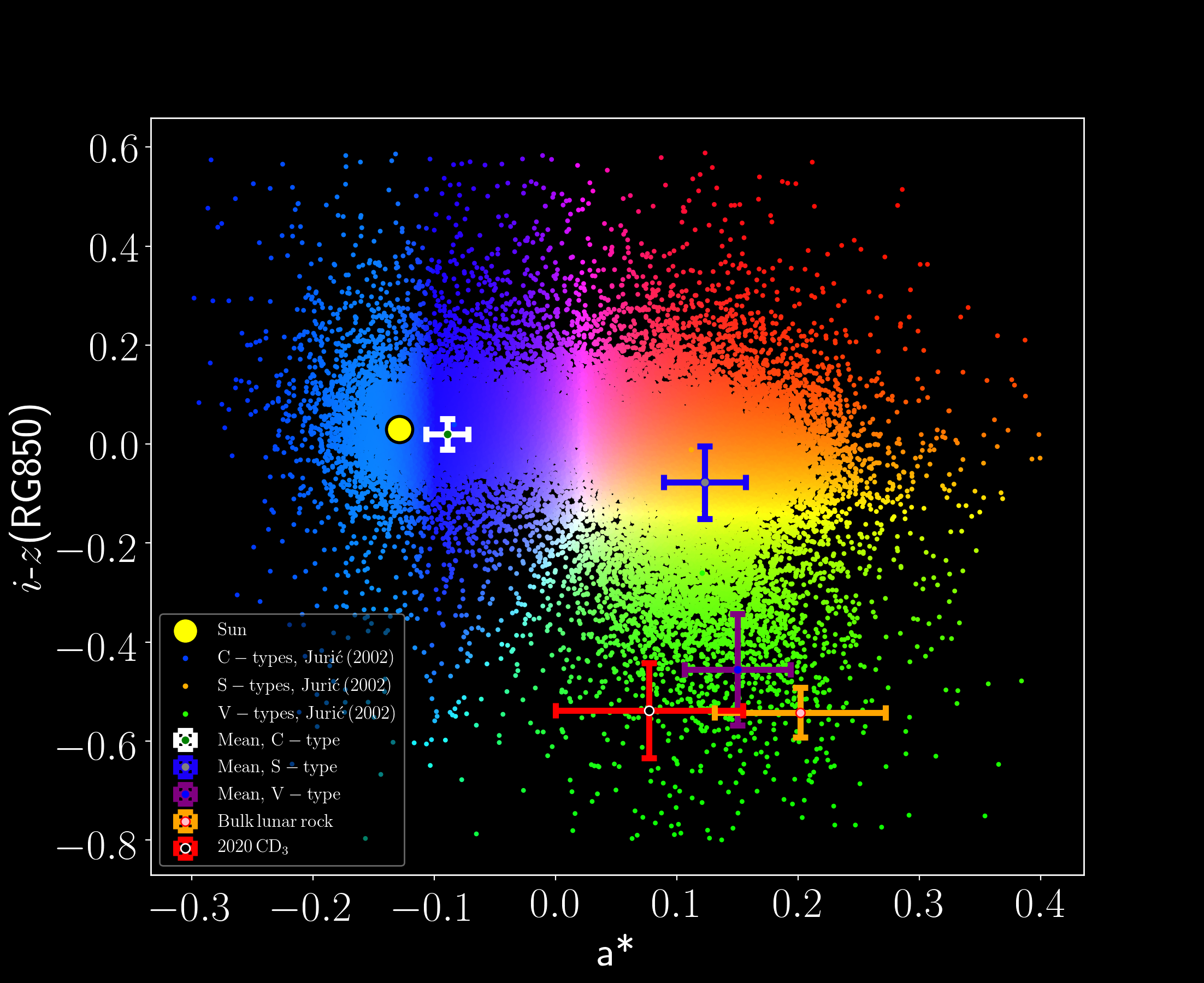}
\caption{a$^{*}$ vs. $i$-$z$(RG850) colors of \cd plotted with a$^{*}$ vs. $i$-$z$ colors of C, S and V type asteroids \citep{Ivezic2001,Juric2002}, active comets \citep[][]{Solontoi2012} and Kuiper Belt Objects \citep[][]{Ofek2012}. The colorization scheme of data points as a function of a$^{*}$ and $i$-$z$ is adapted from \citet[][]{Ivezic2002} where blue colors correspond to C-type asteroids, red colors correspond to S-type asteroids and green colors correspond to V-type asteroids. We note that in this case the measured  RG850 magnitude of \cd is plotted as a substitute for its $z$ magnitude.}
\label{fFiig:colors}
\end{figure}

The colors of \cd are $B$-$V$ = 0.90 $\pm$ 0.07, $V$-$g$ = -0.27 $\pm$ 0.07, $V$-$R$ = 0.46 $\pm$ 0.08, $R$-$I$ = 0.44 $\pm$ 0.06. The equivalent colors in SDSS bands are $g$-$r$ = 0.62 $\pm$ 0.08, $r$-$i$ = 0.21 $\pm$ 0.06 using the filter transformations from \citet[][]{Jordi2006}. In addition, we determine an $i$-RG850 color of -0.54 $\pm$ 0.10. The $B$-$I$ color of \cd is 1.80 $\pm$ 0.05 corresponding to a reflective spectral slope between 434 nm and 761 nm of 18$\pm3\%$/100 nm, indicating a surface significantly redder than the Sun \citep[$B$-$I$ = 1.33, ][]{Holmberg2006aa}. In addition, the parameter a$^{*}$ = (0.89 ($g$-$r$)) + (0.45 ($r$-$i$)) - 0.57, which is an indicator of reflective spectral slope \citep[][]{Ivezic2001}, is equal to 0.08$\pm$0.08 and is plotted vs. $i$-$z$(RG850) in Fig.~\ref{fFiig:colors}. Compared to other asteroids, the a$^{*}$ = 0.08$\pm$0.08 and $i$-$z$(RG850) = -0.54 $\pm$ 0.10 has broad overlap with other V-type asteroids which have on average a$^{*}$ = 0.15 $\pm$ 0.11 and $i$-$z$ = -0.46 $\pm$ 0.04,  \citep[][]{Juric2002}.

\begin{figure}
\centering
\includegraphics[scale=.38]{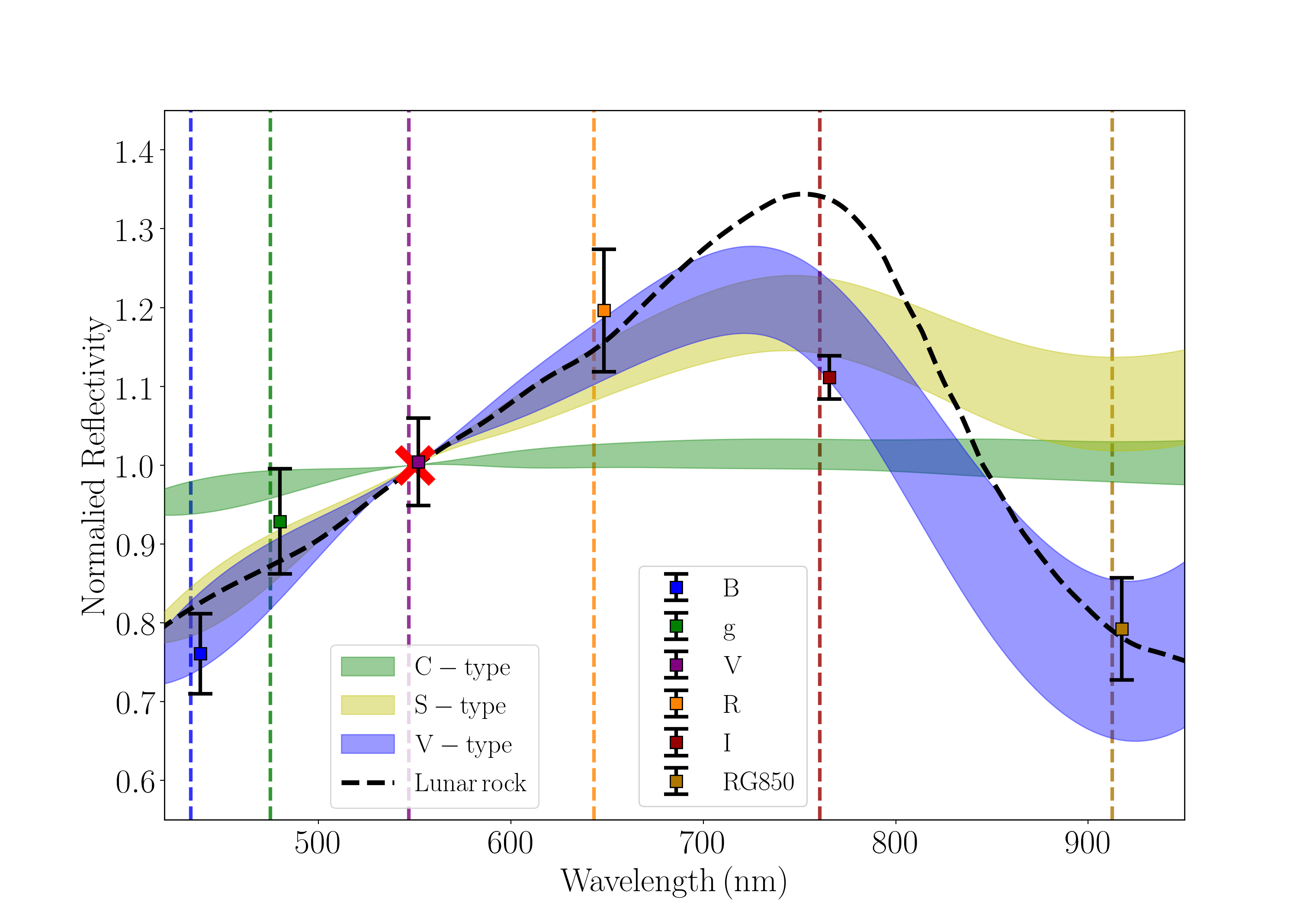}
\caption{Reflectance photometric spectrum of \cd consisting of  $B$, $g$, $V$, $R$, $I$ and RG850 observations of \cd on 2020 March 23 UTC. The $\lambda_{\mathrm{eff}}$ locations of the $B$, $g$, $V$, $R$, $I$ and RG850 filters have been plotted as vertical dashed lines. The data points for the normalized reflectivity of \cd have been offset slightly from their location in the wavelength direction. The error bars on the spectrum data points correspond to 1$\sigma$ uncertainty. The spectrum has been normalized to unity at 550 nm indicated by the red cross. The spectral range of S, V and C-type asteroids from the Bus-DeMeo asteroid taxonomic catalog \citep[][]{DeMeo2009} are over-plotted with the V-type spectrum most closely resembling the spectra of \cd. The average spectrum of coarse bulk Lunar rock samples is plotted for reference \citep[][]{Isaacson2011}.}
\label{fFiig:spectrum}
\end{figure}

To compute the reflectivity spectrum of \cdns, we divide the flux per $B$, $g$, $V$, $R$, $I$ and RG850 filter obtained for \cd by the flux of the Solar analog flux in each corresponding filter. We then normalize the reflectivity spectrum of to 550 nm and detrend the data using a fit of the spectrum with the function
\begin{equation}
r = 1 + a\,(\lambda - 550\, \mathrm{nm})
\end{equation} 
from \citet{Bus2002} where $r$ is the normalized reflectivity as a function of $\lambda$ and $a$ is the spectral slope. The fit used to detrend the data from this function is made by fitting all $B$, $g$, $V$, $R$, $I$ and RG850 data between 435 nm and 912 nm. The resulting normalized reflectivity spectrum is plotted in Fig.~\ref{fFiig:spectrum}. The normalized reflectivity spectrum of \cd is most similar to the spectral range of V-type asteroids \citep[][]{Bus2002, DeMeo2009} with a red slope between 430 nm and 760 nm deep absorption feature in the vicinity of the RG850 data point at $\sim$1000 nm compared to the reflectivity spectra of S and C-type asteroids \citep[][]{DeMeo2009}. We note the same similarity in a$^{*}$ vs. $i$-$z$(RG850) colors between \cd and V-type asteroids as seen in Fig.~\ref{fFiig:colors} as with our normalized reflectivity spectrum. This absorption feature at $\sim$1000 nm as seen for basaltic V-type asteroids found through the inner Main Belt is due to the presence of large amounts of pyroxene minerals on the asteroid's surface \citep[][]{Moskovitz2008}. In addition, we note the same similarity between the spectrum of \cd and bulk basaltic Lunar rock consisting of pyroxenes minerals \citep[][]{Isaacson2011}.

\subsection{Lightcurve, periodicity and axial ratio estimation}
\label{s.lightcurve}

\begin{figure}
\centering
\includegraphics[scale=0.42]{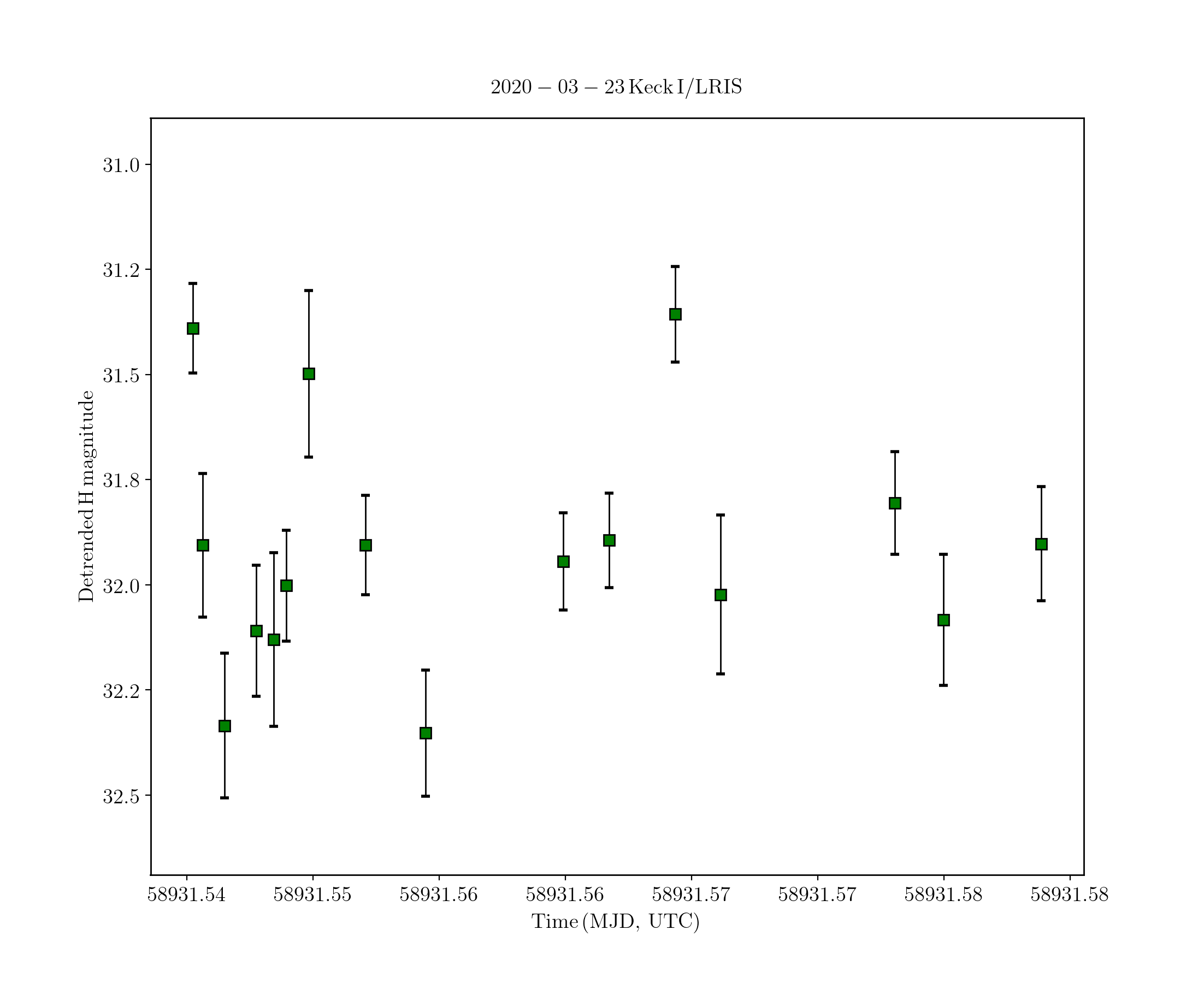}
\includegraphics[scale=0.42]{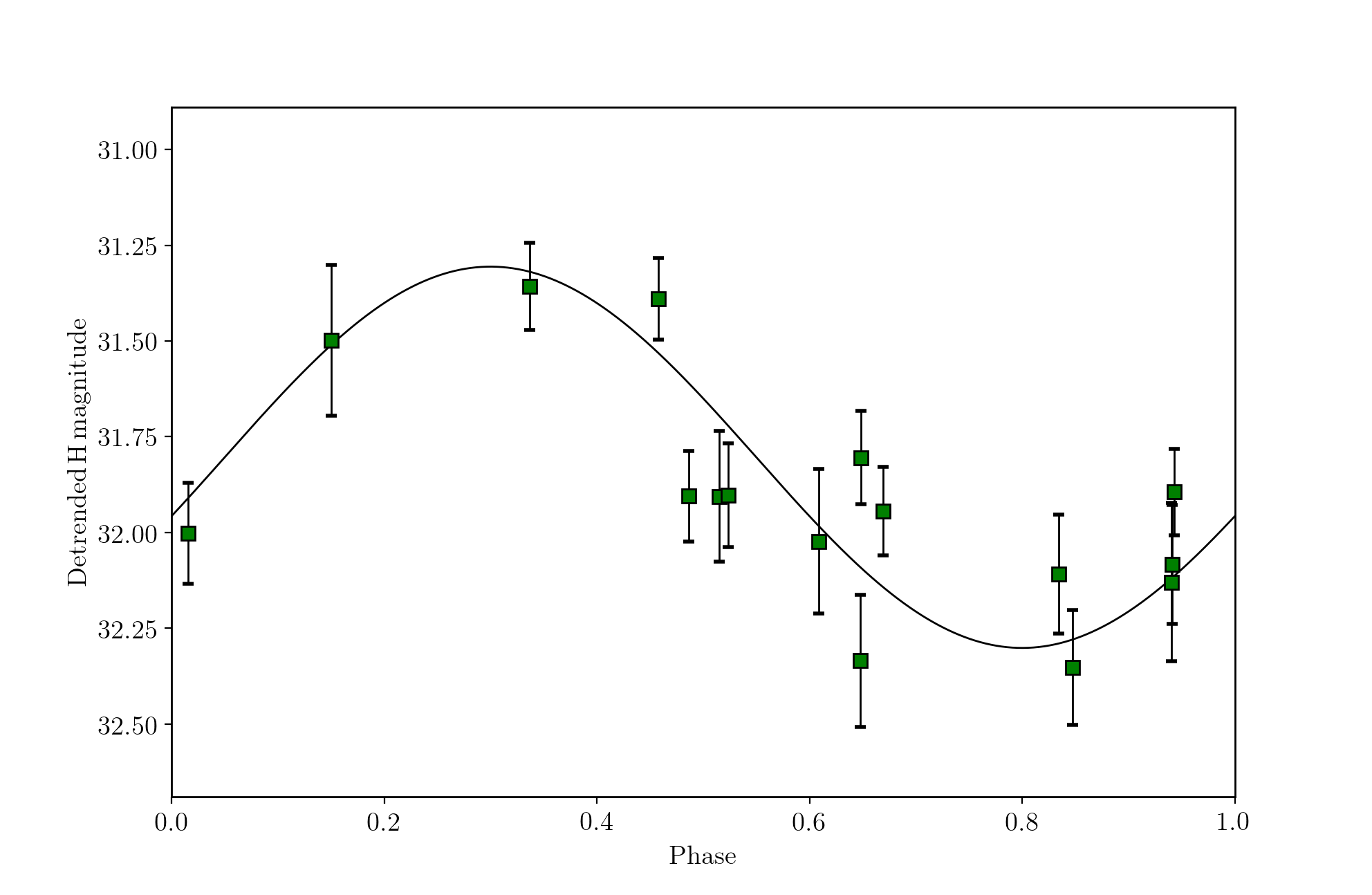}
\caption{Top panel: Detrended H-magntiude lightcurve from 2020 March 23 UTC KecK I/LRIS $B$, $g$, $V$ and $R$ observations of \cd using a 0.81\arcsec~radius aperture radii. The error bars on the data points are equal to their 1 $\sigma$ photometric uncertainties. The data have been detrended and points affected by trailed background stars have been removed. Bottom panel: phased 2020 March 23 UTC KecK I/LRIS observations corresponding to a single-peak lightcurve period of 573.4 s.}
\label{fFiig:kecklightcurve}
\end{figure}

In addition to measuring the photometry of \cd in the per filter $B$, $g$, $V$, $R$, $I$ and RG850 composite image stacks, we search for lightcurve variations by measuring the photometry in our individual $B$, $g$, $V$ and $R$ filter observations. The measured photometric values in the individual images are presented in Table~\ref{t:photo}. We used the colors measured from our composite image stacks described in Section~\ref{sec:photo} and photo-spectrum to convert our $B$, $g$ and $R$ measurements to their equivalent value in $V$. Using our $V$ magnitudes and the following equation:
\begin{equation}
\label{eqn.brightness}
H = V - 5\, \mathrm{log_{10}}(r_h \Delta) +2.5\,\mathrm{log_{10}}\left[ (1 - G)\,\Phi_1(\alpha) + G\,\Phi_2(\alpha) \right ]
\end{equation}
from \citet[][]{Bowell1988} where $r_h$ is the 1.0059 au heliocentric distance of \cd on 2020 March 23 UTC, $\Delta$ is its geocentric distance of 0.0128 au and $\alpha$ is its phase angle of 45.4063$^\circ$. $G$ is the phase coefficient which we use the value of 0.25, the average value of $G$ for S or Q-type asteroids \citep[][]{Veres2015}. $\Phi_1(\alpha)$ and $\Phi_2(\alpha)$ are the basis functions normalized at $\alpha$ = 0$^\circ$ described in \citep[][]{Bowell1988}. We detrend the values of $H$ inferred from Eq.~\ref{eqn.brightness} dividing them by a linear fit which are plotted in the top panel of Fig.~\ref{fFiig:kecklightcurve} which a median value of $H$ = 31.9$\pm$0.1. The errors on these $H$ measurements may be underestimated in part due to the unknown phase function of \cdns. 

\begin{figure}
\centering
\includegraphics[scale=0.43]{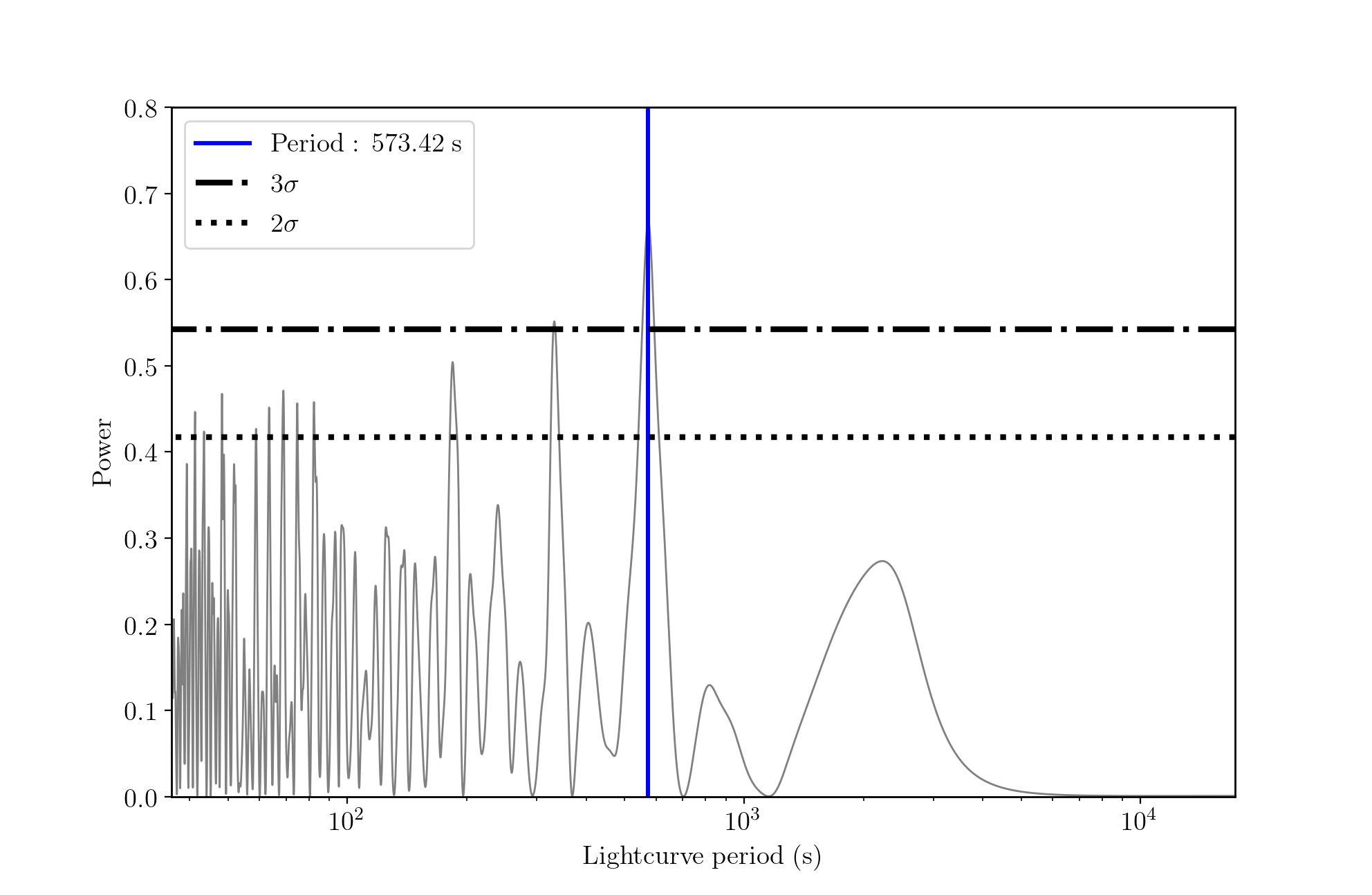}
\includegraphics[scale=0.43]{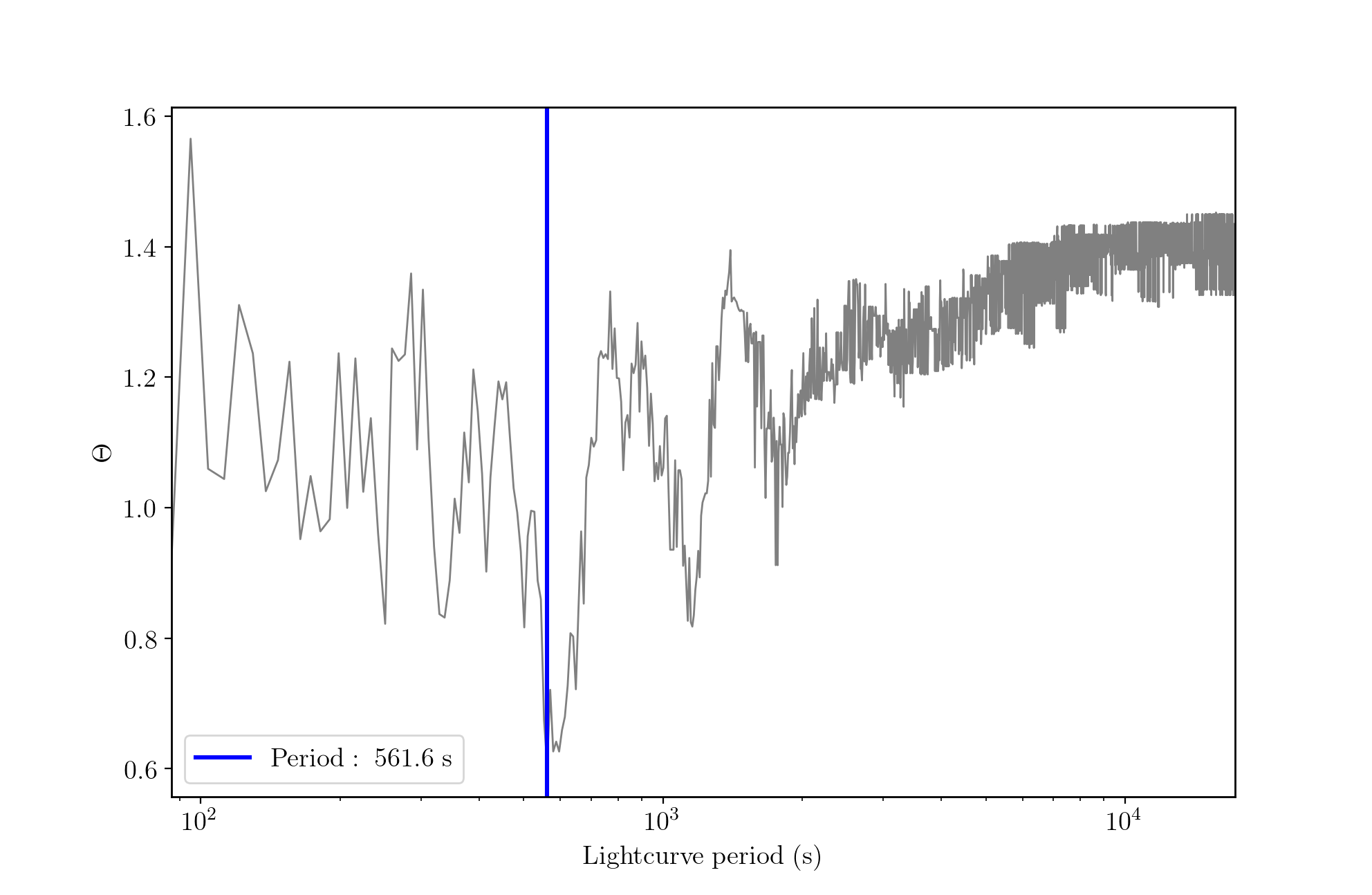}
\caption{Top panel: Lomb-Scargle periodogram of lightcurve period vs. spectral power \citep[][]{Lomb1976} for the Keck I/LRIS lightcurve data from the 2020 March 23 UTC observations. A peak in the power is located at double-peaked lightcurve period of 573.4 s. Bottom panel: Phase dispersion minimization analysis of lightcurve rotation period vs. $\Theta$ metric \citep[][]{Stellingwerf1978}. The $\Theta$ metric is minimized at double-peaked rotation periods of 561.6 s consistent with the 573.4 s rotation period fond with the Lomb-Scargle Periodogram.}
\label{fFiig:periodogram}
\end{figure}

As seen in Fig.~\ref{fFiig:kecklightcurve}, there is brightness variability larger than the $\sim$0.01 photometric scatter measured from the CFHT Skyprobe and the typical $\sim$0.1 mag uncertainty of the data at SNR $\sim$10. Therefore, we will attempt to search for possible periodicities caused by time-variability in \cdns's reflective cross-section over its rotation \citep[][]{Barucci1982}. We apply the Lomb-Scargle periodogram \citep[][]{Lomb1976} to the detrended $H$ magnitude data which is displayed in the top panel of Fig.~\ref{fFiig:periodogram}. Removal of the linear trend over the $\sim$1 h observing period will affect the determination of lightcurve periods that are on $\sim$1 h time scales, but do not affect periodicities on $\sim$100 s time scales. The highest peak in the lightcurve period vs. spectral power curve is located at $\sim$573.4 s with a formal significance of $p \simeq$ 10$^{-4}$. We apply bootstrap estimation \citep[][]{Press1986} of the uncertainties by removing $\sqrt{N}$ data points from the time series lightcurve and repeating our periodogram estimation of the lightcurve period 10,000 times resulting in a central value of $\sim$574.5 s and a 1~$\sigma$ uncertainty estimate of $\sim$30.5 s. As an independent check of our results obtained with the Lomb-Scargle periodogram, we apply phase dispersion minimization analysis to our data \citep[][]{Stellingwerf1978} and obtain a result of $\sim$561.6 s compatible with the lightcurve period estimate obtained with the Lomb-Scargle periodogram as seen in the bottom panel of Fig.~\ref{fFiig:periodogram}. For comparison, the meter-scale asteroids 2006 RH$_{120}$ and 2015 TC$_{25}$ both had lightcurve periods on the order of 60-120 s measured from photometry and radar observations. Furthermore, the ensemble of the available catalog of asteroid lightcurve periods available from the Asteroid Lightcurve Database \citep[][]{Warner2009} seems to indicate that asteroids smaller than 10 m can have rotation periods much shorter than 60 s \citep[][]{Bolin2014}.

We can estimate a rough shape for \cd by assuming it possesses a triaxial prolate shape with dimensions, $a$:$b$:$c$ where $b$  $\geq$ $a$ $\geq$ $c$ in rough approximation with the shapes of other asteroids inferred from lightcurve inversion \citep[][]{Harris2009a, Durech2010}.  Assuming $a$=$c$, the ratio between $b/a$ is described by $b/a \; = \; 10^{0.4 A}$ where $A$ is the peak-to-trough lightcurve amplitude \citep[][]{Binzel1989} resulting in a $b/a$ of $\sim$2.5  for \cd with lightcurve amplitude of $\sim$1 mag. However, the combination of the significant phase angle of $\sim$45$^{\circ}$ \cd was observed on 2020 March 23 UTC and the light scattering properties of its presumably rough surface may have had the effect of amplifying its observed lightcurve amplitude \citep[][]{Zappala1990a}. The relationship between the observed lightcurve amplitude at a given phase angle $\alpha$ and the lightcurve amplitude it would have if observed at $\alpha$ = 0$^{\circ}$ is given by
\begin{equation}
\label{eq.amphase}
\Delta m_{\alpha = 0^{\circ}} = \frac{\Delta m(\alpha)}{1 + s\alpha}
\end{equation}
from \citet[][]{Zappala1990a}, where $s$ is in units of mag deg$^{-1}$ for which we adopt the mean value of 0.012 mag deg$^{-1}$ from light scattering experiment and observations of asteroids \citep[][]{Gutierrez2006}. Applying this correction to our observed lightcurve aplitude of $\sim$1 magnitude lightcurve amplitude at $\sim$45$^{\circ}$ phase angle, we calculate an equivalent $\alpha$ = 0$^{\circ}$ lightcurve amplitude of $\sim$0.6 mag.

In addition to the relationship between phase angle and lightcurve amplitude, the aspect angle of an asteroid when viewed from the Earth can also have an effect on its lightcurve amplitude \citep[e.g.,][]{Hanus2018,Bolin2020HST}. Because the pole orientation of \cd is unknown, it is not possible for us to constrain its aspect viewing angle. Therefore, we adopt the approach of \citet[][]{Bolin2018} which is to average over all possible aspect angles using the following equation
\begin{equation}
\hspace*{-0.75cm}
\label{eq.viewingmag}
  \Delta m_{\rm diff, \theta = 90^\circ} = 1.25\,\mathrm{log}\left( \frac{b^2\cos^2\theta \; + \; c^2\sin^2\theta}{a^2\cos^2\theta \; + \; c^2\sin^2\theta} \right )
\end{equation}
from \citep[][]{Thirouin2016} which gives the lightcurve amplitude of an asteroid when viewed equatorially, i.e., at aspect angle, $\theta$ = 90$^{\circ}$, and where $a$, $b$ and $c$ are the dimensions of \cdns. We will assume 1$\lesssim$$b/a$$\lesssim$2 as observed in asteroid shape models inverted from lightcurves \citep[][]{Hanus2013a,Durech2015} and $a$ = $c$ for a prolate triaxial elipsoid. Integrating Eq.~\ref{eq.viewingmag} over all possible aspect angles results in $\Delta m_{\rm diff}$ $\simeq$ 0.5. Therefore, we calculate the $b/a$ ratio of \cd using $b/a \; = \; 10^{0.4 A}$ where $A$ = ($\Delta m_{\alpha = 0^{\circ}}$ = 0.6 from Eq.~\ref{eq.amphase} + $\Delta m_{\rm diff, \theta = 90^\circ}$ = 0.5 from Eq.~\ref{eq.viewingmag}) $\simeq$ 1 corresponding to a $b/a$ $\sim$2.5 with the lightcurve amplitude phase angle and aspect angle effects roughly canceling each other out.

\subsection{Astrometry, orbit determination and archival data search}
\label{sec:astro}
In addition to measuring the photometry from our observations, we use the positions of \cd measured in our 3 x 30 s $R$ filter images to refine the orbit of \cdns. We measured the astrometry of \cd with the Astrometrica software \citep[][]{Raab2012} combined with reference stars from the \textit{Gaia} data release 2 catalog \citep[][]{Gaia2016,Gaia2018}. Table~\ref{t:astro} contains our measured positions of \cd from our $R$ filter observations. We conservatively estimate an astrometric uncertainty of 1.0\arcsec~in both the right ascension and declination directions to take into account the $\sim$3 s timing uncertainty of the Keck I/LRIS instrument \citep[][]{Burdge2019} resulting in an increased $\sim$0.2\arcsec~uncertainty in the along-track direction measured for the position of \cd in addition to our nominal astrometric uncertainty of $\sim$0.5\arcsec. Adding to our 2020 March 23 UTC observations, we combine our observations with the publicly available observations of \cd measured by other observatories from the Minor Planet Center observation database\footnote{\texttt{https://www.minorplanetcenter.net/tmp/2020\_CD3.txt}} In total, we use 60 observations of \cd taken between 2020 February 18 UTC and 2020 March 23 UTC in addition to our own observations that are listed in Table~\ref{t:astro}. Although uncertainty estimates for other observatories' measurements of asteroids exist \citep[][]{Veres2017}, we adopt conservative estimates for the astrometric uncertainties of $\sim$1.0\arcsec~in both right ascension and declination for these other observatories' measured positions of \cdns. As an exception, we adopt the positional uncertainty of 0.4\arcsec~for the observations for \cd reported by T14, Mauna Kea, UH/Tholen NEO Follow-Up, made by the Canada France Hawaii Telescope and 0.8\arcsec~for observations made by J95, Great Shefford's 0.41 m telescope based on the historical astrometric performance made by these observatories described in the documentation for the orbit fitting software \texttt{Find$\_$Orb} by Bill Gray\footnote{\texttt{https://www.projectpluto.com/find$\_$orb.htm}}. We have submitted our astrometry of \cd on 2020 March 23 UTC to the Minor Planet Center which has appeared in MPEC 2020-O103\footnote{\texttt{https://minorplanetcenter.net/mpec/K20/K20OA3.html}}.

Using \texttt{Find$\_$Orb}, we fit an orbit to our list of observations using the 8 planets and the Moon as perturbers. In addition to the six orbit parameters, semi-major axis, $a$, eccentricity, $e$, inclination, $i$, ascending node, $\Omega$, argument of perihelion, $\omega$, mean anomaly, $M$, we include an additional parameter to our orbital fit, the area-to-mass ration (AMR) as a measure of the effect of Solar radiation pressure on the orbit of \cd \citep[e.g.,][]{Micheli2012}. The nominal orbital fit to our list of observations for the epoch of JD 2,458,931.5 (2020 March 23 UTC) in both heliocentric ($a$, $e$, $i$, $\Omega$, $\omega$, $M$) and geocentric orbital elements ($a_g$, $e_g$, $i_g$, $\Omega_g$, $\omega_g$, $M_g$), the AMR and $H$ magnitude are given in Table~\ref{t:orbit}. It can be noted that the Earth-similar heliocentric elements of $a$$\sim$1 au, $e$$\sim$0.02 and low inclination are typical properties of the minimoon population \citep[][]{Granvik2012, Fedorets2017}.

The mean observed-minus-computed residual from our least squares orbital fit to the observations is 0.40\arcsec~with the Keck I/LRIS observations having observed-minus-computed residuals of $\sim$0.2\arcsec using the seven orbital parameter ($a$, $e$, $i$, $\Omega$, $\omega$, $M$, AMR fit). By comparison, the six orbital paremter ($a$, $e$, $i$, $\Omega$, $\omega$, $M$) fit results in a slightly higher mean observed-minus-computed residual of 0.43\arcsec. The complete list of observed-minus-computed residuals for each of the 60 observations used to compute the orbit is given in Table~\ref{t:astro}.  The $e_g$ of 0.95821 at the epoch of our orbital fit corresponding to the 2020 March 23 UTC data of our observations roughly indicates that \cdns's orbit was approaching a $e_g$$>$1 hyperbolic state for leaving the Earth-Moon system and the measured AMR of 6.96$\pm$2.41$\times$10$^{-4}$ m$^2$/kg is comparable to other small asteroids with measured AMRs \citep[e.g.,][]{Micheli2013,Mommert2014,Farnocchia2017dps}. In addition, \cdns's $i_g$ is retrograde with a value of 146.68615$^{\circ}$ and a geocentric perihelion, $q_g$, of 0.00031 au indicating that it is in the retrograde class of temporary natural satellites that come within the $\sim$0.01 au Hill radius of the Earth \citep[][]{Urrutxua2017,Jedicke2018}.

Our refined orbital solution of \cd from our 2020 March 23 UTC observations enabled the search for possible prediscovery detections of \cd in the Zwicky Transient Facility (ZTF) archive \citep[][]{Masci2019} for additional refinement of the orbit \citep[e.g., as for prediscovery observations interstellar object 2I/Borisov by ZTF][]{Bolin2020asdfasdf,Ye2020AJadf}. The ZTF survey based on the Palomar Observatory's P48 Oschin Schmidt telescope consists of a number of survey programs, some that are open to the public and some that are internal to the ZTF collaboration and Caltech, that are designed to cover the entire sky and detect transient sources including Solar System asteroids and comets \citep[][]{Graham2019}. The ZTF survey camera consists of a 576 megapixel array with a pixel scale of 1.01 arcseconds/pixel covering a 7.4-degree x 7.4-degree field of view \citep[][]{Dekany2016} and $g$, $r$ and $i$ band filters with a $r\sim$20.5 to a SNR = 5 depth in a 30 s exposure generally used in survey \citep[][]{Bellm2019}. The ZTF data system has the ability to detect both round, PSF-like detections \citep[][]{Masci2019} and fast-moving objects moving $>$5\arcsec/m resulting in the detections becoming significantly trailed \citep[][]{Ye2019f, Duev2019} in the survey's 30 s exposures. Therefore, because of its large field of view and ability to identify fast-moving objects, ZTF is the ideal system for ground-based detection of minimoons, objects that typically moving $>$10\arcsec/min or more \citep[][]{Bolin2014,Fedorets2020}.

Extrapolating the trajectory of \cd as far back as 2018 October, we located regions of the sky where it was covered by the ZTF survey in $g$ and $r$ filters. We narrowed our search for prediscovery observations to times when \cd was brighter than $V$$\sim$20 taking into account the $\sim$0.2 mag mean color differences between the standard Johsnon $V$  filter and the ZTF $r$ filter for asteroids \citep[][]{Veres2017}. Our search revealed dates when \cd was brighter than $V$$\sim$20 on 2019 January 17 UTC ($V$$\sim$19.1), 2019 April 04 UTC ($V$$\sim$15.2), 2019 November 15 UTC ($V$$\sim$19.5) and 2020 February 13 UTC ($V$$\sim$17.0). However, the only date overlapping with ZTF observations was on 2019 November 15 UTC in which a single $g$ band exposure was obtained which was also during the full phase of the moon greatly increasing the sky background in the image. In addition, \cd was moving nearly $\sim$40\arcsec/min resulting in significant trailing losses \citep[][]{Shao2014} making its already difficult brightness of $V$$\sim$19.5 impossible to detect. A possible method of detection for \cd is to use where its orbital trajectory overlaps with higher-cadence fields while it is moving with a slower rate of motion and use synthetic tracking to shift and stack along its possible trajectories increasing its detection's SNR to a detectable threshold as has been demonstrated for ZTF data of Main Belt and NEOs \citep[][]{Zhai2020}, however, a full demonstration of synthetic tracking to locate \cd in ZTF data is beyond the scope of this work.

\subsection{Orbital evolution}

The second-known minimoon \cd was discovered while it was captured by the Earth-Moon system. To determine its orbital evolution before, during and after its captured state, we implemented the \texttt{rebound} $n$-body orbit integration package \citep[][]{Rein2012} with our fitted orbit from Table~\ref{t:orbit}. In addition to its nominal orbit, we cloned $\sim$10 additional massless test particles defined from the vertices of a cuboid represented by the heliocentric orbital elements and $\sigma$ orbital parameter semi-major axis $a$, eccentricity $e$ and inclination $i$ uncertainties listed in Table.~\ref{t:orbit} and an initial ephemeris time of 2020 March 23 UTC. The simulations are run using the \texttt{IAS15} integrator \citep[][]{Rein2015} and the Sun, eight major masses of the Solar system, along with the Moon, Vesta, Ceres and Pluto \footnote{Taken from the JPL HORIZONS Solar System Dynamics Database \url{https://ssd.jpl.nasa.gov/} \citep{Giorgini1996JPLSSdatabase}, on 10th April, 2020.}. The simulations were run using a timestep of 0.00249 y (21.825 hours, 0.03 times the Lunar orbital period), with an output of 0.01 y for 5-y and 100-y time-frames.

\begin{figure}
\centering
\includegraphics[scale=.6]{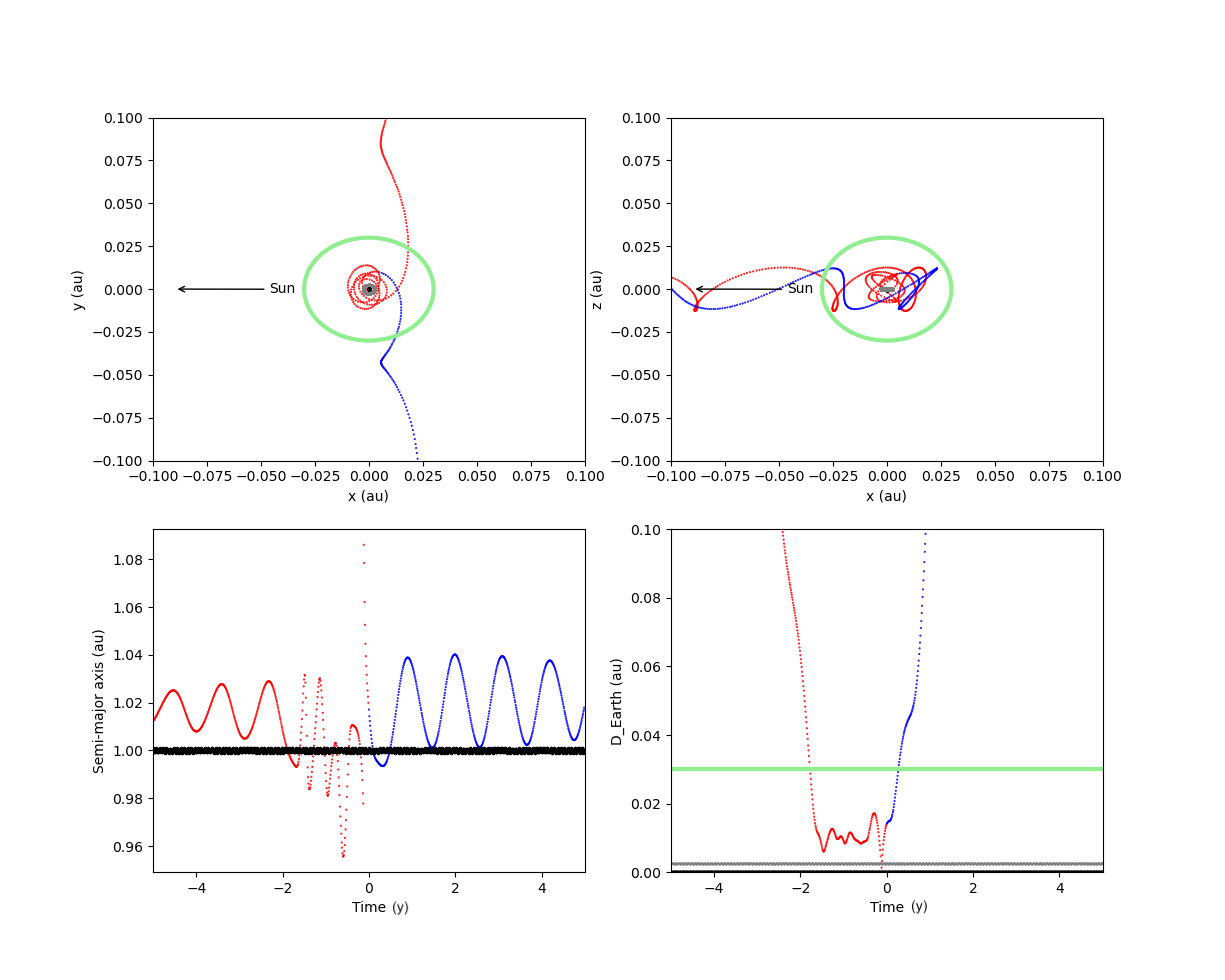}
\caption{Top left panel: mean geocentric, co-rotating Cartesian y and x coordinates of \cd orbital clones $\pm$5 y centered on 2020 March 23 UTC encompassing its $\sim$700 day capture completing $\sim$5 revolutions around the Earth-Moon system. The red dotted line indicates the trajectory of \cd before 2020 March 23 UTC and the blue dotted line indicates the trajectory of \cd after 2020 March 23 UTC. A green circle with a radius of three times the Earth's Hill radii of $\sim$0.03 au is overplotted. The direction towards the Sun in the co-rotating frame is indicated. Top right panel: same as the top left panel except for mean geocentric, co-rotating Cartesian x and z coordinates. Bottom left panel: the evolution of \cdns's orbital clones' mean semi-major axes $\pm$5 y centered on 2020 March 23 UTC. The color code of the dotted lines is the same as in the top panels. Bottom right panel: the mean geocentric distance of \cd orbital clones $\pm$5 y centered on 2020 March 23 UTC. A horizontal green line indicates three times the Hill radii in distance. The color code of the dotted lines is the same as in the previous three panels.}
\label{fFiig:capture}
\end{figure}

We adopt the definition of geocentric capture from \citet[][]{Fedorets2017} and \citet[][]{Jedicke2018} to describe the geocentric orbital evolution of \cdns, namely that while captured, \cd remains within 3 Hill radii ($\sim$0.03 au) of the Earth, has a $e_g$$<$1, and approaches the Earth to within 1 Hill radius ($\sim$0.01 au) at some point during its captures. As seen in Fig.~\ref{fFiig:capture}, \cdns, approaches the Earth-Moon system opposite from the Sun's direction in the direction of the L2 Lagrange point with its capture beginning in mid 2018 with a low $\sim$1 km/s encounter velocity. Almost half of minimoons pass through the L2 Lagrange point while becoming temporarily geocentrically bound \citep[][]{Granvik2012}, therefore, it seems \cdns's capture is non-exceptional in the case of temporarily captured asteroids. In addition, we see from the top panels of Fig.~\ref{fFiig:capture} that \cd is captured on a retrograde orbit $\sim$100$^\circ$ and completes $\sim$5 revolutions around the Earth-Moon system while remaining within three Hill radii of geocenter. Integrating its orbit forward and backward, the majority of \cd orbital clones remained captured within the Earth-Moon for $\sim$2 y as seen in the bottom right panel of Fig.\ref{fFiig:capture} leaving the Earth-Moon system in mid 2020. Integrating the orbit with and without a Solar radiation pressure component does not significantly affect the capture duration of \cdns. The geocentric orbit of \cd is retrograde for nearly the entirety of its capture and its final orbit will result in it having a slightly larger heliocentric semi-major axis of $a$ of 1.027 au compared to its pre-capture $a$ of 0.973 au as seen in the bottom left panel of Fig.~\ref{fFiig:capture}. Overall it seems the capture of \cd is a typical, however, having a  longer duration than the $\sim$1 y capture duration of 2006 RH$_{120}$, the only other known minimoon, and the $\sim$9 month capture duration averaged over the minimoon population \citep[][]{Granvik2012}.

\begin{figure}
\centering
\includegraphics[scale=.6]{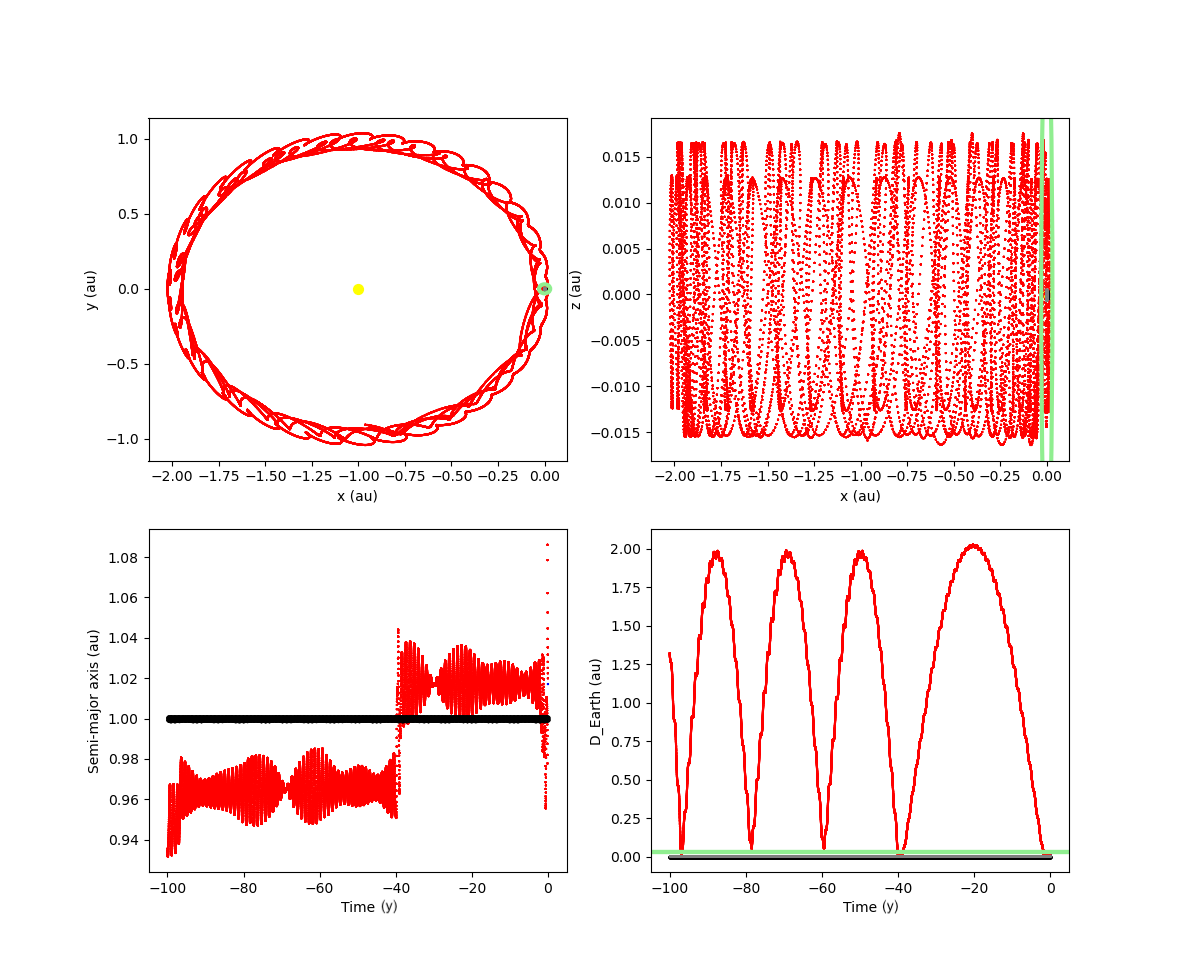}
\caption{Top left panel: mean geocentric Cartesian y and x coordinates of \cd orbital clones integrated backwards 100 y from 2020 March 23 UTC (blue line) with the Earth's three Hill radii marked in green. Top right panel: same as the top left panel except for the mean geocentric Cartesian x and z coordinates of \cd orbital clones integrated backwards 100 y from 2020 March 23 UTC. Bottom left panel: the evolution in \cdns's orbital clones' mean semi-major axis integrated backwards 100 y from 2020 March 23 UTC with the Earth's orbit in black. Bottom right panel: the geocentric distance of \cd orbital clones integrated backwards 100 y from 2020 March 23 UTC.}
\label{fFiig:backwards}
\end{figure}

\begin{figure}
\centering
\includegraphics[scale=.6]{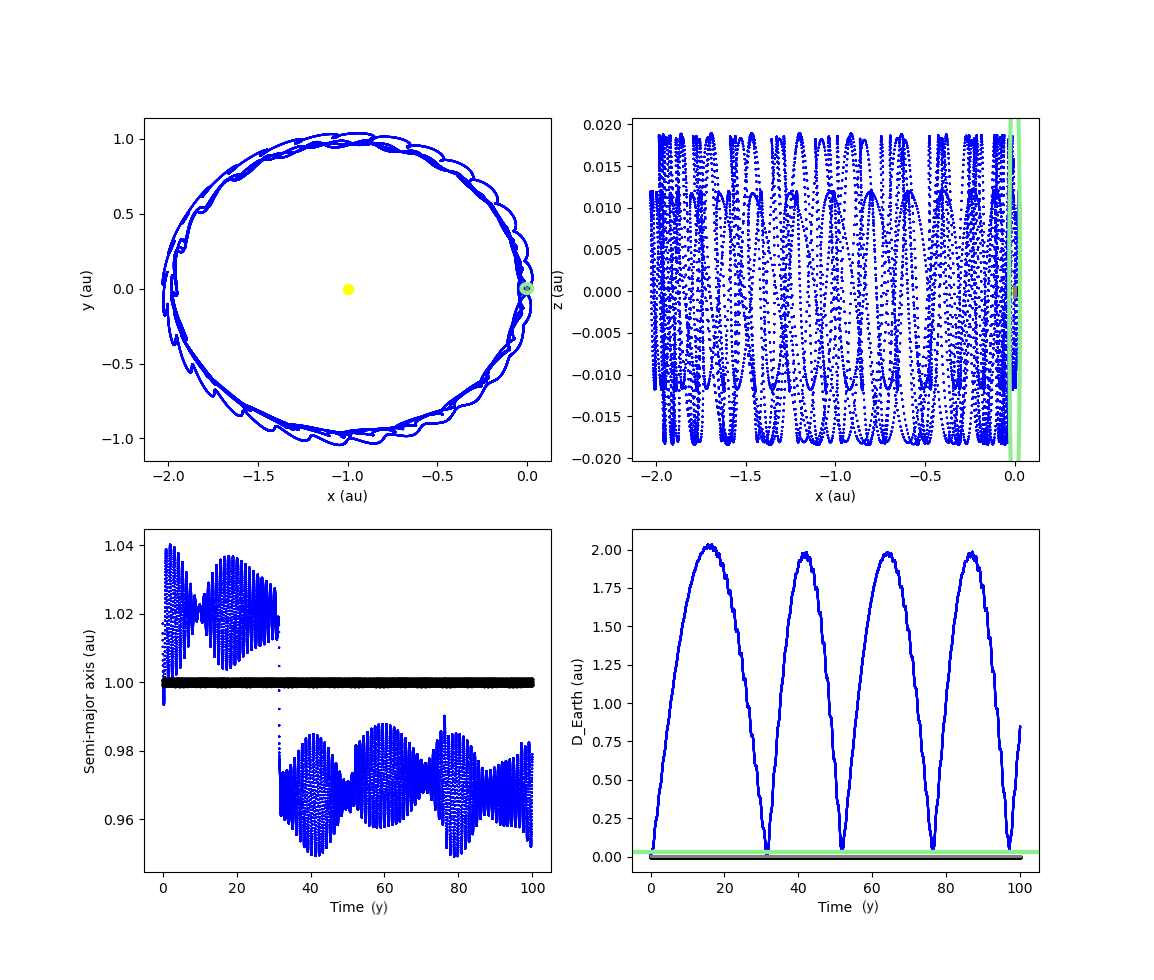}
\caption{Top left panel: same as in Fig.~\ref{fFiig:backwards} except for orbital clones of \cd integrated forwards 100 y from 2020 March 23 UTC (red line) with the Earth's three Hill radii marked in green. Top right panel: same as the top left panel except for the mean geocentric Cartesian x and z coordinates of \cd orbital clones integrated forwards 100 y from 2020 March 23 UTC. Bottom left panel: the evolution in \cdns's orbital clones' mean semi-major axis integrated forwards 100 y from 2020 March 23 UTC with the Earth's orbit in black. Bottom right panel: the geocentric distance of \cd orbital clones integrated forwards 100 y from 2020 March 23 UTC.}
\label{fFiig:forwards}
\end{figure}

In addition, we take a look at the longer term, 100 y heliocentric orbital evolution of \cd as presented in Figs.~\ref{fFiig:backwards} and \ref{fFiig:forwards}. Integrating the orbit of \cd 100 y into the past and into the future show similar behavior in that \cd has close encounters with the Earth placing \cd inside the Hill radius of the Earth every $\sim$20-30 y as seen in the bottom right panels of Figs.~\ref{fFiig:backwards} and \ref{fFiig:forwards}. The long-term orbit of \cd resembles a horseshoe orbit as seen in the upper left panels of Figs.~\ref{fFiig:backwards} and \ref{fFiig:forwards} where its status as temporarily capture asteroids has resulted from its similar orbital plane and low encounter velocity relative to Earth's \citep[][]{Granvik2013b,Jedicke2018}. Interestingly, some of the \cd orbit clones when integrated into the future switch from a max inclination of 0.012$^\circ$ to 0.031$^\circ$ during the next encounter with the Earth. In addition, we have undertaken preliminary, long term simulations using the hybrid MERCURIUS \texttt{rebound} integrator, \citep{Rein2019MERCURIUS} using the same initial conditions as above, including the eight clones. These initial simulations indicate that the horse-shoe dynamical situation is stable for at least $\sim$10$^6$ years.

\section{Discussion and Conclusions}

It appears that \cd represents a typical case when compared to the known ensemble and dynamical path of minimoons. While it is difficult to estimate the true population of minimoons given the vast incompleteness of asteroid surveys \citep[][]{Jedicke2016} the discovery of \cd along with 2006 RH$_{120}$ confirm minimoons as viable members of the near-Earth object population and is the first minimoon to be spectrophotometrically characterized. It seems as its orbital dynamics are similar to co-orbitals of Earth \citep[][]{Morais2002} of which there is one known example, 2010 TK$_7$ \citep[][]{Connors2011} and quasi-satellites of which several are known, e.g., (469219) 2016 HO$_3$ \citep[][]{Chodas2016}. Compared to the asteroid population at large, out of $\sim$1 million known asteroids as of 2020 July, only $\sim$10 are of similar size as \cd with H$\sim$31 or smaller making \cd one of the smallest asteroids discovered and characterizes with spectrophotometry \citep[e.g.,][]{Reddy2016}.

While its spectrum and colors seem to indicated that \cd is a likely V-type asteroid with an origin from the inner-Main Belt \citep[][]{DeMeo2013} as discussed in Section~\ref{sec:photo}, we can use its orbit in reference with models describing the NEO population \citep[][]{Granvik2016, Granvik2018} as an independent indication of its source through asteroid escape pathways in the Main Belt \citep[][]{Granvik2017}. Comparison with the NEO population model suggests that its most likely Main Belt escape source with $\sim$70$\%$ probability was through the $\nu_6$ resonance located near the inner edge of the Main Belt at 2.1 au for low inclination objects \citep[][]{Milani1990}. The second and third most likely sources are the Hungaria asteroid population located between 1.8 au and 2.0 au \citep[][]{Milani2010b} with a $\sim$25$\%$ probability and the 3:1 mean motion resonance located at the border between the inner and center Main Belt at 2.5 au \citep[][]{Wisdom1983} with a $\sim$5$\%$ probability. We note that the current NEO model is only available for asteroids with $H$ 25 or brighter, therefore we have made the comparison between \cd and the NEO model with the assumption that it has $H$ = 25. Its other orbital parameters remain the same for the purposes of comparison with the NEO model.

Weighing the NEO albedo model \citep[][]{Morbidelli2020} according to these source probabilities for \cd results in a predicted albedo, $p_v$, of $\sim$0.23 which is on the lower 0.25-0.45 $p_v$ range of V-type asteroids \citep[][]{DeMeo2013}. Using our measured $H$ magnitude of 31.9$\pm$0.1 and the following equation relating the diameter $D$ and $p_v$:
\begin{equation}
D = \frac{1329}{\sqrt{p_v}}10^{-\frac{H}{5}}
\end{equation}
from \citet[][]{Harris2002}, we calculate that \cd has $D$ = 1.2$\pm$0.1 m assuming $p_v$ = 0.23 as determined for \cd from its possible Main Belt sources in the NEO albedo model \citep[][]{Morbidelli2020} or  $D$ = 0.9$\pm$0.1 m if using $p_v$ = 0.35, the mean albedo of V-type asteroids in the Main Belt \citep[][]{DeMeo2013} making it currently the smallest asteroid studied spectrophometrically with the next smallest being asteroid 2015 TC$_{25}$ \citep[][]{Reddy2016}.

We estimate the density of \cd by combining our constraints on its diameter and albedo combined with our AMR measurement from fitting its orbit.  Using our measured AMR of 6.9$\pm$2.4$\times$10$^{-4}$ m$^2$/kg and diameter estimate of $D$ = 1.0$\pm$0.1 m, we estimate the bulk density of \cd to be 2.1$\pm$0.7 g/cm$^3$ broadly compatible with the densities of other small asteroids determined from AMR measurements  \citep[e.g.,][]{Micheli2012}. In comparison, the density of 2015 TC$_{25}$ is $\sim$1 g/cm$^3$ assuming a diameter of $\sim$2.2 m \citep[][]{Reddy2016} and an AMR of 6-7$\times$10$^{-4}$ m$^2$/kg \citep[][]{Farnocchia2017dps}. Assuming 0$\%$ macroporosity, the total mass of \cd is $\sim$10$^4$ kg.

While its estimated density of 2.1$\pm$0.7 g/cm$^3$ is broadly consistent with the density of V-type asteroids which have bulk densities of $\sim$2.3 g/cm$^3$ \citep[][]{Carry2012} which its spectrum resembles, it is likely, however, that \cd has a porosity in the range of $\sim$10$\%$-20$\%$ as for meteorites \citep[][]{Consolmagno2008}, its closest analog as one of the smallest known asteroids. In contrast, the km-scale V-type asteroids which we are drawing in comparison with \cd have macroporosities of $\sim$30$\%$ or larger \citep[][]{Carry2012} resulting in a higher density when correcting their $\sim$2.3 g/cm$^3$ bulk densities for their higher macroporosity. Therefore, it may be more appropriate to compare the density of \cd with achondritic basaltic meteorites which typically have bulk densities of $\sim$3.0 g/cm$^3$ which is somewhat larger than our estimated range of the density of \cdns. 

\cd is likely the product of the fragmentation of a larger parent asteroid given its small $\sim$1 m size and its correspondingly short $<$1 Myr time scale \citep[][]{Bottke2005b}. There appears to be some discrepancy in the fact that \cd most likely originates in the 1.8-2.2 au range at the inner edge of the Main Belt while having a spectrum similar to V-type asteroids which are thought to originate from the asteroid (4) Vesta \citep[][]{Binzel1993,Parker2008} and are primarily located at $\sim$2.3 au \citep[][]{DeMeo2013}. A significant number of V-type asteroids exist further from the Sun located past 2.5 au in the central Main Belt \citep[][]{Carruba2005,Migliorini2017} which could provide a possible source of Earth-crossing V-type NEOs like \cd if they were to drift inward into the 3:1 resonance due to the thermal recoil Yarkovsky effect \citep[][]{Farinella1998}. 

However,  the role of the 3:1 resonance in transporting \cd into Earth-crossing space from the Main Belt seems unlikely due to its $\sim$5$\%$ source probability. In addition, the Yarkovsky effect is able to transport meter-scale objects like \cd into the proximity of the $\nu_6$ resonance, the most likely source of \cdns, in $<$1 Myrs if it were to have originated as a fragment at 2.3 au \citep[][]{Bottke2006, Vokrouhlicky2015} where most V-type asteroids are found even if possesses significantly different thermal inertia properties compared to larger, km-scale asteroids \citep[][]{Delbo2007,Bolin2017b}. In addition, is apparent from the wide distribution of $\lesssim$1 km Vesta family fragments covering the entirety of the inner Main Belt \citep[][]{Bolin2017} that the size-dependent velocity distribution of family fragments originating from Vesta could have placed \cdns-sized objects anywhere between the $nu_6$ resonance at $\sim$2.2 au and the 3:1 resonance at 3.5 au \citep[][]{Carruba2016d,Bolin2017a}. Therefore, the location of the $\nu_6$ resonance at the inner edge of the Main Belt at 2.2 au as the most likely source of \cd does not necessarily preclude asteroids far from its vicinity as the original parent body of \cdns.

Besides collisions, rotational fission of asteroids that are spinning near their rotational stability limit could be a possible origin of \cd \citep[][]{Walsh2008}. Several asteroids have been observed to be in the act of rotationally shedding mass or fragmenting \citep[e.g.,][]{Moreno2017,Jewitt2017,Ye2019adfs} or have a dynamically-associated cluster of asteroids compatible with a fragmentation event in the recent past due to their rotation \citep[][]{Vokrouhlicky2017a}. In addition, binary asteroids systems can become decoupled over time due to the influence of thermal radiation recoil effects \citep[][]{McMahon2010} which can result in small asteroids like \cd leaving their binary systems and entering Earth-crossing space. The fragmentation of asteroid parent bodies or decoupling of binary systems can occur while an asteroid parent body in near-Earth object space \citep[][]{Scheirich2019,Bottke2020adfads} providing an origin for \cd outside of the Main Belt.

Another possible origin of minimoons is from Lunar impacts. While the orbits of Lunar debris dynamically decay after a few kyrs, it is possible that some Lunar ejecta can be re-captured by the Earth-Moon system as minimoons due to their orbital similarity with the Earth \citep[][]{Gladman1995}. As presented in Fig.~\ref{fFiig:spectrum}, the spectrum of \cd is compatible with the spectrum of bulk Lunar rock at the precision of our spectrophotometry.  In addition, our inferred density of \cd of 2.1$\pm$0.7 g/cm$^3$ is similar impact basin ejecta Lunar rock which have bulk density \citep[$\sim$2.4 g/cm$^3$][]{Kiefer2012} and $\sim$20$\%$ porosity. Under the assumption that \cd originated as Lunar ejecta, the young, $\lesssim$1 Myr-scale cosmic-ray exposure ages of Lunar meteorites \citep[][]{Eugster2006} implies that the vast majority of Lunar meteorites and \cd by extension had to have been produced by a large and recent Lunar impact. The most recent, large impact that could produce ejecta the size of \cd is the Giordano Bruno crater that has been estimated to be $\sim$4 Myrs old \citep[][]{Morota2009} based on the occurrence of craters near its proximity. However, the Lunar ejecta origin of \cd is diminished by the fact that the vast majority Lunar meteorites posses cosmic ray exposure ages much shorter than the 4 Myr age of the Giordano Bruno crater suggesting that the dominant source of recent Lunar meteorites and thus Lunar ejecta are much smaller, more recent impact events than could have produced ejecta the size of \cd \citep[][]{Minton2019}.

\acknowledgments

This work was supported by the GROWTH project funded by the National Science Foundation under PIRE Grant No 1545949.

Some of the data presented herein were obtained at the W. M. Keck Observatory, which is operated as a scientific partnership among the California Institute of Technology, the University of California and the National Aeronautics and Space Administration. The Observatory was made possible by the generous financial support of the W. M. Keck Foundation.

The authors wish to recognize and acknowledge the very significant cultural role and reverence that the summit of Maunakea has always had within the indigenous Hawaiian community. We are most fortunate to have the opportunity to conduct observations from this mountain.

C.F.~gratefully acknowledges the support of his research by the Heising-Simons Foundation ($\#$2018-0907).

M.~W.~Coughlin acknowledges support from the National Science Foundation with grant number PHY-2010970. 

Based on observations obtained with the Samuel Oschin Telescope 48-inch and the 60-inch Telescope at the Palomar Observatory as part of the Zwicky Transient Facility project. ZTF is supported by the National Science Foundation under Grant No. AST-1440341 and a collaboration including Caltech, IPAC, the Weizmann Institute for Science, the Oskar Klein Center at Stockholm University, the University of Maryland, the University of Washington, Deutsches Elektronen-Synchrotron and Humboldt University, Los Alamos National Laboratories, the TANGO Consortium of Taiwan, the University of Wisconsin at Milwaukee, and Lawrence Berkeley National Laboratories. Operations are conducted by COO, IPAC, and UW.

This work has made use of data from the European Space Agency (ESA) mission \textit{Gaia} (\texttt{https://www.cosmos.esa.int/gaia}), processed by the \textit{Gaia} Data Processing and Analysis Consortium (DPAC, \texttt{https://www.cosmos.esa.int/web/gaia/dpac/\\consortium}). Funding for the DPAC has been provided by national institutions, in particular the institutions participating in the \textit{Gaia} Multilateral Agreement.

\facility{Keck I Telescope, P48 Oschin Schmidt telescope/Zwicky Transient Facility} 

\bibliographystyle{aasjournal}
\bibliography{ms}

\begin{longtable}{|c|c|c|c|}
\caption{Summary of \cd photometry taken on 2020 March 23 UTC.\label{t.photometry1}}\\
\hline
Date$^1$ & Filter$^2$ & Exp$^3$&$H^4$ \\
(MJD UTC)&&(s)&\\
\hline
\endfirsthead
\multicolumn{4}{c}%
{\tablename\ \thetable\ -- \textit{Continued from previous page}} \\
\hline
Date$^1$ & Filter$^2$ & Exp$^3$&$H^4$ \\
UTC&&(s)&\\
\hline
\endhead
\hline \multicolumn{4}{r}{\textit{Continued on next page}} \\
\endfoot
\hline
\endlastfoot
58931.5452386 & R & 60 s & 31.39 $\pm$ 0.11\\
58931.5456206 & g & 120 s & 31.91 $\pm$ 0.17\\
58931.5465002 & R & 60 s & 32.33 $\pm$ 0.17\\
58931.5477386 & R & 60 s & 32.11 $\pm$ 0.16\\
58931.5484447 & g & 120 s & 32.13 $\pm$ 0.21\\
58931.5489424 & R & 30 s & 32.00 $\pm$ 0.13\\
58931.5498336 & R & 30 s & 31.50 $\pm$ 0.2\\
58931.5520674 & B & 120 s & 31.90 $\pm$ 0.12\\
58931.5544632 & B & 120 s & 32.35 $\pm$ 0.15\\
58931.5599146 & B & 120 s & 31.94 $\pm$ 0.12\\
58931.5617317 & B & 120 s & 31.89 $\pm$ 0.11\\
58931.5643474 & g & 120 s & 31.36 $\pm$ 0.11\\
58931.566153 & g & 120 s & 32.02 $\pm$ 0.19\\
58931.5730511 & V & 120 s & 31.80 $\pm$ 0.12\\
58931.5749956 & V & 120 s & 32.08 $\pm$ 0.16\\
58931.5788613 & V & 120 s & 31.90 $\pm$ 0.14\\
\hline
\caption{Columns: (1) observation date correct for light travel time; (2) Keck I/LRIS Filter; (3) Exposure time (4) $V$ band equivalent $H$ magnitude with 1 $\sigma$ uncertainties}
\label{t:photo}
\end{longtable}

\newpage
\begin{longtable}{|c|c|c|c|c|c|c|c|}
\caption{Summary of astrometry from observations taken by Keck I/LRIS and other observatories between 2020 February 15 UTC and 2020 March 23 UTC.\label{t.astrometry}}\\
\hline
Date$^1$ & R.A$^2$ & Dec.$^3$&$\sigma_{\mathrm{R.A.}}$$^4$&$\sigma_{\mathrm{Dec.}}$$^5$&$X_{res.}$$^6$&$Y_{res.}$$^7$&Obs. code$^8$\\
(UTC)&&&(\arcsec)&(\arcsec)&(\arcsec)&(\arcsec)&\\
\hline
\endfirsthead
\multicolumn{4}{c}%
{\tablename\ \thetable\ -- \textit{Continued from previous page}} \\
\hline
Date$^1$ & R.A$^2$ & Dec.$^3$&$\sigma_{\mathrm{R.A.}}$$^4$&$\sigma_{\mathrm{Dec.}}$$^5$&$X_{res.}$$^6$&$Y_{res.}$$^7$&Obs. code$^8$\\
(UTC)&&&(\arcsec)&(\arcsec)&(\arcsec)&(\arcsec)&\\
\hline
\endhead
\hline \multicolumn{4}{r}{\textit{Continued on next page}} \\
\endfoot
\hline
\endlastfoot
2020 Feb 15.511140 & 13 03 33.110 & +09 10 43.10 & 1.00 & 1.00 & -0.33 & -0.02 & G96\\
2020 Feb 15.516240 & 13 03 34.520 & +09 13 03.60 & 1.00 & 1.00 & +0.07 & +0.59 & G96\\
2020 Feb 15.521330 & 13 03 35.960 & +09 15 21.90 & 1.00 & 1.00 & -0.50 & +0.47 & G96\\
2020 Feb 15.545470 & 13 03 44.540 & +09 26 01.30 & 1.00 & 1.00 & -0.17 & -0.78 & G96\\
2020 Feb 15.545640 & 13 03 44.640 & +09 26 05.90 & 1.00 & 1.00 & +0.27 & -0.60 & G96\\
2020 Feb 15.545820 & 13 03 44.700 & +09 26 11.10 & 1.00 & 1.00 & +0.06 & -0.08 & G96\\
2020 Feb 15.545990 & 13 03 44.770 & +09 26 15.40 & 1.00 & 1.00 & +0.05 & -0.20 & G96\\
2020 Feb 15.995517 & 13 18 35.410 &+12 12 02.10 & 1.00 & 1.00 & -0.36 & +0.15 & L01\\
2020 Feb 15.997407&13 18 36.130 &+12 12 45.40 & 1.00 & 1.00 & +0.05 & +0.38 & L01\\
2020 Feb 15.999831&13 18 36.990 &+12 13 40.90 & 1.00 & 1.00 & +0.02 & +0.80 & L01\\
2020 Feb 16.004013&13 18 38.390 &+12 15 14.70 & 1.00 & 1.00 & -0.30 & -0.03 & L01\\
2020 Feb 16.005639&13 18 38.940 &+12 15 51.70 & 1.00 & 1.00 & -0.02 & +0.31 & L01\\
2020 Feb 16.363950 &13 25 20.450 &+14 16 52.80 & 1.00 & 1.00 & +0.26 & +0.52 & 291\\
2020 Feb 16.366530 &13 25 20.430 &+14 17 41.60 & 1.00 & 1.00 & +0.17 & +0.61 & 291\\
2020 Feb 16.369160 &13 25 20.380 &+14 18 30.70 & 1.00 & 1.00 & +0.09 & +0.26 & 291\\
2020 Feb 16.437910 &13 25 08.050 &+14 38 18.00 & 1.00 & 1.00 & +0.14 & +0.05 & I52\\
2020 Feb 16.439840 &13 25 07.630 &+14 38 49.70 & 1.00 & 1.00 & +0.02 & +0.18 & I52\\
2020 Feb 16.441770 &13 25 07.230 &+14 39 20.80 & 1.00 & 1.00 & +0.25 & -0.15 & I52\\
2020 Feb 16.443700 &13 25 06.810 &+14 39 52.00 & 1.00 & 1.00 & +0.23 & -0.25 & I52\\
2020 Feb 17.040534&13 34 23.090 &+16 43 52.20 & 1.00 & 1.00 & +1.20 & +0.32 & L01\\
2020 Feb 17.046622&13 34 22.810 &+16 45 14.60 & 1.00 & 1.00 & -0.06 & -0.01 & L01\\
2020 Feb 17.051069&13 34 43.300 &+16 40 45.90 & 0.80 & 0.80 & -0.21 & -0.02 & J95\\
2020 Feb 17.054297&13 34 22.440 &+16 46 58.00 & 1.00 & 1.00 & +0.04 & +0.44 & L01\\
2020 Feb 17.070077&13 34 45.600 &+16 45 07.30 & 0.80 & 0.80 & +0.06 & -0.33 & J95\\
2020 Feb 17.084942&13 34 46.670 &+16 48 26.90 & 0.80 & 0.80 & -0.13 & -0.01 & J95\\
2020 Feb 17.100726&13 34 47.270 &+16 51 52.70 & 0.80 & 0.80 & +0.07 & -0.36 & J95\\
2020 Feb 17.508000 &13 38 02.170 &+18 18 19.10 & 1.00 & 1.00 & +0.37 & +0.61 & G96\\
2020 Feb 17.510030 &13 38 01.600 &+18 18 36.30 & 1.00 & 1.00 & +0.82 & +0.12 & G96\\
2020 Feb 17.512060 &13 38 00.990 &+18 18 54.50 & 1.00 & 1.00 & +0.60 & +0.76 & G96\\
2020 Feb 17.514090 &13 38 00.380 &+18 19 11.50 & 1.00 & 1.00 & +0.28 & +0.32 & G96\\
2020 Feb 17.980051&13 45 25.930&+19 19 39.42 & 1.00 & 1.00 & -0.89 & +0.35 & Z84\\
2020 Feb 17.998253&13 45 29.233&+19 23 16.64 & 1.00 & 1.00 & -0.80 & -0.04 & Z84\\
2020 Feb 18.016457&13 45 30.942&+19 26 49.53 & 1.00 & 1.00 & -0.79 & -0.11 & Z84\\
2020 Feb 18.414650 &13 47 52.480 &+20 25 18.00 & 1.00 & 1.00 & +0.09 & +0.40 & G96\\
2020 Feb 18.415230 &13 47 52.300 &+20 25 23.20 & 1.00 & 1.00 & +0.17 & +0.51 & G96\\
2020 Feb 18.415810 &13 47 52.140 &+20 25 27.20 & 1.00 & 1.00 & +0.54 & -0.57 & G96\\
2020 Feb 18.416390 &13 47 51.930 &+20 25 33.00 & 1.00 & 1.00 & +0.21 & +0.16 & G96\\
2020 Feb 19.301670 &13 55 16.820 &+21 55 58.10 & 1.00 & 1.00 & -0.51 & +0.82 & G96\\
2020 Feb 19.303920 &13 55 16.910 &+21 56 17.50 & 1.00 & 1.00 & +0.23 & -0.70 & G96\\
2020 Feb 19.306180 &13 55 16.960 &+21 56 39.00 & 1.00 & 1.00 & +0.71 & -0.15 & G96\\
2020 Feb 20.418810 &14 00 29.620 &+23 35 56.90 & 1.00 & 1.00 & -0.02 & +0.59 & G96\\
2020 Feb 20.419560 &14 00 29.390 &+23 36 00.70 & 1.00 & 1.00 & +0.44 & -0.01 & G96\\
2020 Feb 20.420320 &14 00 29.140 &+23 36 05.20 & 1.00 & 1.00 & +0.69 & +0.05 & G96\\
2020 Feb 20.421070 &14 00 28.880 &+23 36 11.00 & 1.00 & 1.00 & +0.76 & +0.97 & G96\\
2020 Feb 21.095901&14 03 49.976&+24 21 03.47 & 1.00 & 1.00 & -0.39 & -0.35 & Z84\\
2020 Feb 21.120492&14 03 42.513&+24 23 16.74 & 1.00 & 1.00 & -0.47 & -0.68 & Z84\\
2020 Feb 21.146268&14 03 33.897&+24 25 23.50 & 1.00 & 1.00 & +0.03 & -0.04 & Z84\\
2020 Feb 21.172268&14 03 24.721&+24 27 15.50 & 1.00 & 1.00 & +0.09 & -0.23 & Z84\\
2020 Feb 21.196581&14 03 16.142&+24 28 46.33 & 1.00 & 1.00 & +0.54 & -0.46 & Z84\\
2020 Feb 27.642330&14 19 58.612&+29 10 39.85 & 0.40 & 0.40 & +0.00 & +0.01 & T14\\
2020 Feb 27.643742&14 19 58.040&+29 10 40.82 & 0.40 & 0.40 & -0.01 & +0.01 & T14\\
2020 Feb 27.645156&14 19 57.468&+29 10 41.76 & 0.40 & 0.40 & -0.04 & +0.02 & T14\\
2020 Mar 01.477510 &14 23 44.930 &+30 15 25.40 & 1.00 & 1.00 & +0.40 & -0.33 & G37\\
2020 Mar 01.479210 &14 23 44.270 &+30 15 27.40 & 1.00 & 1.00 & -0.36 & -0.14 & G37\\
2020 Mar 01.480250 &14 23 43.970 &+30 15 28.30 & 1.00 & 1.00 & +0.52 & -0.32 & G37\\
2020 Mar 22.483420&14 21 40.530&+33 16 25.12 & 0.40 & 0.40 & -0.04 & +0.00 & T14\\
2020 Mar 22.487379&14 21 38.999&+33 16 27.56 & 0.40 & 0.40 & +0.03 & +0.01 & T14\\
2020 Mar 23.544957&14 20 00.300 &+33 15 49.70 & 1.00 & 1.00 & +0.03 & +0.01 & 568\\
2020 Mar 23.548839&14 19 58.780 &+33 15 48.50 & 1.00 & 1.00 & -0.10 & -0.13 & 568\\
2020 Mar 23.549729&14 19 58.420 &+33 15 47.90 & 1.00 & 1.00 & +0.28 & +0.13 & 568\\
\hline
\caption{Columns: (1) UTC observation date at the midpoint of the exposure; (2) right ascension; (3) declination; (4) uncertainty in right ascension; (5) uncertainty in declination; (6) observed-minus-computed residual in the $X$ direction; (7) observed-minus-computed residual in the $Y$ direction; (8) Minor Planet Center Observatory Code}
\label{t:astro}
\end{longtable}

\newpage
\begin{table}
\centering
\caption{Orbital elements of 2020 AV$_2$ based on observations collected between 2020 February 15 UTC and 2020 March 23 UTC. The orbital elements are shown for the Julian date (JD) shown using the software \texttt{Find$\_$Orb} by Bill Gray. The 1~$\sigma$ uncertainties are given in parentheses.}
\label{t:orbit}
\begin{tabular}{ll}
\hline
Heliocentric Elements&
\\ \hline
Epoch (JD) & 2,458,931.5\\
\hline
Time of perihelion, $T_p$ (JD) & 2,458,907.045$\pm$(0.019)\\
Semi-major axis, $a$ (au) & 1.01713$\pm$(6.49x10$^{-7}$)\\
Eccentricity, $e$ & 0.02858$\pm$(6.58x10$^{-7}$)\\
Perihelion, $q$ (au) & 0.98806$\pm$(6.51x10$^{-8}$)\\
Aphelion, $Q$ (au) & 1.04620$\pm$(1.34x10$^{-6}$)\\
Inclination, $i$ ($^{\circ}$) & 0.55483$\pm$(1.80x10$^{-5}$)\\
Ascending node, $\Omega$ ($^{\circ}$) & 116.954$\pm$(1.70x10$^{-3}$)\\
Argument of perihelion, $\omega$ ($^{\circ}$) & 357.557$\pm$(1.30x10$^{-3}$)\\
Mean Anomaly, $M$ ($^{\circ}$) & 65.1632$\pm$(4.44x10$^{-4}$)\\
\hline
Geocentric Elements&\\
\hline
Epoch (JD) & 2,458,931.5\\
\hline
Time of perihelion, $T_{p,g}$ (JD) & 2,458,893.615$\pm$(7.47x10$^{-5}$)\\
Semi-major axis, $a_g$ (au) & 0.00752$\pm$(3.72x10$^{-7}$)\\
Eccentricity, $e_g$ & 0.95821$\pm$(9.88x10$^{-7}$)\\
Perihelion, $q_g$ (au) & 0.00031$\pm$(2.05x10$^{-8}$)\\
Aphelion, $Q_g$ (au) & 0.01472$\pm$(7.24x10$^{-7}$)\\
Inclination, $i_g$ ($^{\circ}$) & 146.68615$\pm$(7.24x10$^{-7}$)\\
Ascending node, $\Omega_g$ ($^{\circ}$) & 309.888$\pm$(5.5x10$^{-4}$)\\
Argument of perihelion, $\omega_g$ ($^{\circ}$) & 280.529$\pm$(1.0x10$^{-3}$)\\
Mean Anomaly, $M_g$ ($^{\circ}$) & 99.299$\pm$(7.0x10$^{-3}$)\\
\hline
Area-to-Mass ratio, AMR (m$^2$/kg)& 6.96x10$^{-4}$$\pm$(2.41x10$^{-4}$)\\
Absolute Magnitude, $H$ & 31.9$\pm$(0.1)\\
\hline
\end{tabular}
\end{table}

\end{document}